\documentclass[%
 reprint,
 amsmath,amssymb,
 prd,
floatfix,
]{revtex4-2}

\def\babar{\mbox{\slshape B\kern-0.1em{\smaller A}\kern-0.1em
    B\kern-0.1em{\smaller A\kern-0.2em R}}}

\def\Bbar {\kern 0.18em\overline{\kern -0.18em B}{}\xspace}

\def\keVsquared{keV/$c^{2}$}

\newcommand{\mevc}{\ensuremath{{\mathrm{\,Me\kern -0.1em V\!/}c}}\xspace}
\newcommand{\gevcc}{\ensuremath{{\mathrm{\,Ge\kern -0.1em V\!/}c^2}}\xspace}
\newcommand{\mevcc}{\ensuremath{{\mathrm{\,Me\kern -0.1em V\!/}c^2}}\xspace}

\usepackage[pdfencoding=auto, psdextra]{hyperref}
\usepackage{bookmark}
\usepackage[export]{adjustbox}
\usepackage{relsize}
\usepackage{graphicx}
\usepackage{dcolumn}
\usepackage{bm}
\usepackage{graphicx}
\usepackage{xspace}
\usepackage[caption=false]{subfig}
\usepackage{rotating}
\usepackage[]{hyperref}
\usepackage[utf8]{inputenc}
\usepackage[dvipsnames]{xcolor}
\hypersetup{
colorlinks,
linkcolor={blue},
linktocpage=true,
citecolor={red},
urlcolor={olive}
}

\begin{document}
\babar-PUB-22/002 \\
\vspace{0.3cm} 
 \hspace{0.15cm} SLAC-PUB-17695 \\

\title{Search for Heavy Neutral Leptons Using Tau Lepton Decays at \babar}

\author{J.~P.~Lees}
\author{V.~Poireau}
\author{V.~Tisserand}
\author{E.~Grauges}
\author{A.~Palano}
\author{G.~Eigen}
\author{D.~N.~Brown}
\author{Yu.~G.~Kolomensky}
\author{M.~Fritsch}
\author{H.~Koch}
% ...
\collaboration{The \babar\ Collaboration}
\noaffiliation

\begin{abstract}
This article presents a model independent search for an additional, mostly sterile, Heavy Neutral Lepton (HNL), that is capable of mixing with the Standard Model $\tau$ neutrino with a mixing strength of $|U_{\tau 4}|^{2}$, corresponding to the absolute square of the extended Pontecorvo-Maki-Nakagawa-Sakata (PMNS) matrix element. Data from the \babar~  experiment, with a total integrated luminosity of $424~ \text{fb}^{-1}$, are analyzed using a kinematic approach that makes no assumptions on the model behind the origins of the HNL, its lifetime or decay modes. No significant signal is found. Upper limits on $|U_{\tau 4}|^{2}$ at the 95$\%$ confidence level, depend on the HNL mass  hypothesis and vary from $2.31 \times 10^{-2}$ to  $5.04 \times 10^{-6}$ (with all uncertainties considered), across the mass range $100 < m_{4} < 1300 ~\text{MeV}/c^{2}$; the more stringent limits being placed at higher masses. 
\end{abstract}

\maketitle


\section{\label{sec:motivations}Motivations}
Heavy Neutral Leptons (HNLs) are predicted by many extensions of the Standard Model (SM) to explain several phenomena. They interact via gravity but have no electric charge, no weak hypercharge, no weak isospin, and no color charge; HNLs have no ordinary weak interactions, except those induced by mixing. They are generally considered singlets under all gauge interactions and are often referred to as ``sterile neutrinos." A theoretical overview and experimental review of recent searches for HNLs can be found in  Refs. \cite{PBC,Abdullahi:2022jlv}.

Observation of neutrino oscillations has established the non-zero mass of at least two of the SM neutrinos. Absolute values of these masses are yet to be determined, but experiments have measured the mass squared differences, with current bounds detailed in Ref.\cite{newpdg}. Sterile neutrinos have long been used to explain the apparent smallness of the SM neutrino masses \cite{small}. 

Heavy Neutral Leptons could also be responsible for the generation of the matter-antimatter asymmetry of the Universe \cite{matter_anti} via leptogenesis \cite{baryo,lepto1}. Leptogenesis scenarios can predict HNLs at a mass scale as low as $\mathcal{O}$(\gevcc)\cite{lepto2}, thus these theories can be explored in current, and near-future, particle physics experiments. 

The neutrino Minimal Standard Model (or $\nu$-MSM)\cite{neut_min} is one theory that predicts HNLs at the GeV/$c^{2}$-scale. In $\nu$-MSM adding three sterile, right-handed, Majorana HNLs to the SM can explain neutrino oscillations, the origin of the baryon asymmetry of the Universe \cite{DM2,astro}, and provide a dark matter candidate \cite{DM1}. Two of these HNLs have masses in the \mevcc~ to \gevcc range and a third, the dark matter candidate, has mass at the \keVsquared-scale. The $\nu$-MSM is compatible with all current measurements.

Sterile fermions of masses $\mathcal{O}(\text{eV}/c^{2})$ can also explain the anomalies in very short baseline oscillation measurements and cosmological data analyses \cite{sterile_lowmass}. Recent re-analysis of data from the GALLEX \cite{Kaether:2010ag} and SAGE \cite{galium} solar neutrino experiments has exposed an unexplained 14$\pm$5$\%$ deficit in the number of recorded $\nu_{e}$; referred to as the ``gallium anomaly." In addition, numerous analyses of the flux of $\bar{\nu}_{e}$ from reactors have suggested a deficit of $\bar{\nu}_{e}$ at the 98.6$\%$ confidence level (C.L.) \cite{reactor}; denoted as the ``reactor anti-neutrino anomaly." A third anomaly  --- the ``accelerator anomaly" --- stems from measurements at the LSND \cite{PhysRevD.64.112007} experiment that evaluated the oscillation $\nu_{e} \rightarrow \nu_{\mu}$ at a baseline of $L$ = 30m. The LSND experiment measured an excess of neutrinos at the level of 3.8$\sigma$, which could be explained by the existence of a sterile neutrino with a mass $\mathcal{O}(\text{eV}/c^{2})$. Further support for this excess was presented by the MiniBooNE experiment at the 2.8$\sigma$ level \cite{Mboone}, although recent results from MicroBooNE do not yet support this anomaly \cite{micro}.

Heavy Neutral Leptons with mass in the \mevcc~ to \gevcc range could be produced in $\tau$ decays, giving rise to deviations from the SM expectations. In this article the possibility of an additional neutrino state (associated with the HNL) interacting with the $\tau$-lepton, via charged-current weak interactions, is considered. Mixing between the HNL mass eigenstate and the active neutrinos can be parameterized by the extended Pontecorvo–Maki–Nakagawa–Sakata (PMNS) matrix with additional elements $U_{l4}$,  where $l$ denotes the SM lepton flavor state i.e. $e,\mu, \tau$:

\begin{equation*}
    \begin{pmatrix}
    \nu_{e}\\
    \nu_{\mu}\\
    \nu_{\tau}\\
    \nu_{\text{\tiny L}}
    \end{pmatrix}
    =
    \begin{pmatrix}
    U_{e1} &  U_{e2} &  U_{e3} &  U_{e4}\\
    U_{\mu 1} &  U_{\mu 2} &  U_{\mu 3} &  U_{\mu 4}\\
    U_{\tau 1} &  U_{\tau 2} &  U_{\tau 3} &  U_{\tau 4}\\
    U_{\text{\tiny L} 1} & U_{\text{\tiny L} 2} & U_{\text{\tiny L} 3} & U_{\text{\tiny L} 4} \\
    \end{pmatrix}
    \begin{pmatrix}
    \nu_{1}\\
    \nu_{2}\\
    \nu_{3}\\
    \nu_{4}
    \end{pmatrix},
\end{equation*}
where $L$ represents some hypothetical additional lepton flavor. Analyses of cosmological data and $Z$ boson decays, summarized in Ref. \cite{newpdg}, are consistent with there being only three charged lepton flavors. Here, the PMNS matrix is extended for just one HNL, but others can be added in the same way. The PMNS matrix for the anti-neutrinos is identical to that for neutrinos under CPT symmetry.  Experimental data on the mixing strength between the $\tau$-sector and a HNL are limited. Although the probability of a fourth neutrino state interacting with the electron ($|U_{e4}|^{2}$) or muon ($|U_{\mu4}|^{2}$) has tight constraints \cite{newpdg}, limits on $|U_{\tau 4}|^{2}$ are weaker, motivating the possibility that $|U_{\tau 4}| \gg |U_{e 4}|$, $|U_{\mu 4}|$. In this article a search for a HNL with mass in the range 100 \mevcc $< m_{4} <$ 1300 \mevcc is presented. Existing bounds in the range $\sim$ 300 \mevcc~ to $\sim$ 1 \gevcc~ range are particularly weak.

\section{Current Bounds}

Numerous experiments have searched for the existence of HNLs with mass from $\mathcal{O}$(eV/$c^2$) up to hundreds of GeV/$c^2$, with no evidence seen. 

Robust bounds on the mixing of heavy neutrinos with both electron and muon neutrinos have been provided by searches for excesses in the missing mass distribution of pions and kaons leptonic decays. Strong constraints on couplings have already been set by several experiments, such as: PS191 \cite{PS191}, CHARM \cite{CHARM}, NuTeV \cite{NuTeV}, E949 \cite{E949}, PIENU \cite{PIENU}, TRIUMF-248 \cite{tri}, NA3 \cite{NA3} and NA62 \cite{NA62:2021bji}. 

The current limits on $|U_{\tau 4}|^{2}$ come from the  NOMAD ~\cite{Nomad}, CHARM~\cite{CHARM} and DELPHI~\cite{DELPHI} experiments. More recently ArgoNeuT published limits \cite{PhysRevLett.127.121801} in the 280 $-$ 970 \mevcc range through a search for a Dirac HNL decaying to $\nu \mu^{+} \mu^{-}$. The CHARM experiment conducted a search for HNLs produced in the decay of neutral particles into two electrons and provided limits on $|U_{\tau 4}|^{2}$ in the mass range 10 $-$ 290 \mevcc. The NOMAD experiment collected $4.1 \times 10^{19}$ 450 GeV protons on target at the WANF facility at CERN. A search for $D_{s} \rightarrow \tau \nu_{R}$ followed by $\nu_{R} \rightarrow \nu_{\tau}  e^{+}  e^{-}$ was conducted, resulting in an upper limit on  $|U_{\tau 4}|^{2}$ in the mass range from 10 $-$ 190 \mevcc. The DELPHI experiment at LEP provided limits on $|U_{\tau 4}|^{2}$ in the \gevcc~ mass range through searching for signatures of HNLs decaying to visible SM particles, specifically $e^{+}e^{-} \rightarrow Z \rightarrow \nu \nu_{4}$. In Ref.~\cite{mesons} the $\tau$ and meson branching ratios are used to further constrain the parameter space.

\section{\label{sec:experiment}Data Sample and Detector}

The data sample used in this analysis corresponds to an integrated luminosity of 424 fb$^{-1}$ \cite{BaBar:2013agn}, collected with the ~\babar~
  detector at the PEP-II $e^{+}e^{-}$ storage ring at the SLAC
  National Accelerator Laboratory. At PEP-II, 9 GeV electrons collide with 3.1 GeV positrons at center-of-mass (CM) energies near 10.58 GeV, on the $\Upsilon(4S)$ resonance; 10$\%$ of the data were recorded 40 MeV below this resonance. The average cross-section for $\tau$-pair production of electron-positron annihilation is $ \sigma (e^{+}e^{-} \rightarrow \tau^{+} \tau^{-}) =(0.919 \pm 0.003) $ nb \cite{newpdg}; thus the data sample corresponds to $\sim 4 \times 10^{8}$ produced $\tau$-pairs, before applying any reconstruction or selection criteria.

In the \babar~ detector~\cite{AUBERT20021,BaBar} a Silicon Vertex Tracker (SVT) and a 40 layer Drift Chamber (DCH), placed inside a 1.5-T solenoid magnet, are utilized to reconstruct charged-particle tracks. The transverse momentum resolution is $0.47\%$ at 1 GeV/$c$, where the transverse momentum, $p_{T}$, is defined as the total momentum of all four tracks orthogonal to the beam axis.

An Electro-Magnetic Calorimeter (EMC) measures the energy of electrons and photons with a resolution of $3\%$ at 1 GeV. A Ring Imaging Cherenkov detector (DIRC) is located in front of the EMC and is used, together with specific ionization loss measurements in the SVT and DCH, to identify charged  pions and kaons, and provide additional electron identification. The instrumented flux return of the solenoid is used to identify muons.

\section{Experimental Strategy}
\label{exp_strat}

The analysis approach in this study was originally proposed in Ref. \cite{proposal} and follows from that used by ALEPH to attempt to determine the $\tau$ neutrino mass~\cite{ALEPH}. We consider that HNL can interact with the tau via charged-current weak interactions. The key principle is that if the decay products of the $\tau$ have recoiled against a heavy neutrino, the phase space and the kinematics
of the visible particles would be modified with respect to SM $\tau$ decay with a massless neutrino. We assume that the HNL does not decay within the detector.

This search studies the 3-prong, pionic $\tau$ decay since it has a relatively large branching fraction and gives access to the region 300$ <m_{4}< $1360 \mevcc~ (up to the kinematic endpoint), which historically has weaker constraints, whilst pion channels of higher multiplicity would only test the lower mass region that is already well constrained. It should be noted that the mass of the SM $\tau$ neutrino is unknown and the current upper limit on the heaviest neutrino is $<$18.2 (95$\%$ confidence level (C.L.)) \mevcc~ \cite{newpdg}. In this analysis all SM neutrinos are assumed to have zero mass; changing the SM neutrino masses from 0 to the experimental upper limit induces negligible changes in the kinematic distributions used.

The 3-prong decay can be considered a 2 body decay: 

\begin{equation}  
\tau^{-} \rightarrow h^{-} (E_{h}, \vec{p}_{h}) + \nu  (E_{\nu}, \vec{p}_{\nu}), 
\end{equation}
where $h^{-}$ denotes the hadronic system and $\nu$ describes the outgoing neutrino state. An analogous equation could be written for the $\tau^{+}$ channel. The allowed phase space of the reconstructed energy, $E_{h}$, and invariant-mass, $m_{h}$, of the hadronic system would vary as functions of the mass of the HNL. As the HNL gets heavier the proportion of the original $\tau$-lepton's energy going to the visible pions diminishes.
 
In the CM frame the $\tau$-lepton energy is assumed to be $\sqrt{s}/2$, when there is no initial state radiation. Since the direction of the decaying $\tau$-lepton is not known we cannot compute the neutrino mass directly but we know that $E_{h}$ must fall between two extremes that define the kinematically allowed values:

\begin{equation} E_{\tau} - \sqrt{m^{2}_{4} + q_{+}^{2}} < E_{h} < E_{\tau} - \sqrt{m^{2}_{4} + q_{-}^{2}}, \end{equation}
where
\begin{widetext}
\begin{equation}
q_{\pm} = \frac{m_{\tau}}{2} \bigg ( \frac{m_{h}^{2} - m_{\tau}^{2} - m_{4}^{2}}{m_{\tau}^{2}} \bigg ) \sqrt{\frac{E_{\tau}^{2}}{m_{\tau}^{2}} - 1} \pm \frac{E_{\tau}}{2} \sqrt{\big ( 1- \frac{(m_{h}+ m_{4})^{2}}{m_{\tau}^{2}}\big ) \big ( 1- \frac{(m_{h} - m_{4})^{2}}{m_{\tau}^{2}} \big )};
\end{equation}
\end{widetext}
and $3m_{\pi^{\pm}} < m_{h} < m_{\tau} - m_{4}$. As the HNL mass increases, the allowed phase space of the visible system is reduced in the $E_{h}-m_{h}$ plane.

The observed kinematic phase space distributions of the hadronic system could be assumed to be a superposition of two phase spaces: the one associated with the heavy neutrino (weighted by $|U_{\tau 4}|^{2}$), and that associated with a decay to the SM  neutrino (weighted by (1 -  $|U_{\tau 4}|^{2}$)). For a hypothetical mixing with the $\tau$-lepton, the total differential decay rate would then be:
\begin{widetext}
\begin{equation} \frac{\text{d}\Gamma(\tau^{-} \rightarrow \nu h^{-})}{\text{d}m_{h}dE_{h}}\bigg|_{\text{Total}} = |U_{\tau 4}|^{2} \frac{\text{d}\Gamma(\tau^{-} \rightarrow \nu h^{-})}{\text{d}m_{h}\text{d}E_{h}}\bigg|_{\text{HNL}} +   (1-  |U_{\tau 4}|^{2}) \frac{\text{d}\Gamma(\tau^{-} \rightarrow \nu h^{-})}{\text{d}m_{h}\text{d}E_{h}}\bigg|_{\text{SM}}.   \end{equation} 
\end{widetext}

In this analysis we search for a HNL signal by comparing the observed event yield density in the ($m_{h},E_{h}$) plane to a set of template 2D histogram distributions for the backgrounds, obtained by simulating all $\tau$ known decays as well as non-$\tau$ background events, and the potential HNL signal for different $m_{4}$ mass values. Although the invariant-mass and outgoing hadronic energy ($m_{h}$ and $E_{h}$) are correlated, more information can be extracted by considering both variables simultaneously.  

At the \babar~ collision energy of $\sqrt{s} = 10.58$ GeV, the process of $e^{+}e^{-} \rightarrow \tau^{+} \tau^{-}$ produces $\tau$ leptons that have decays well-separated in the CM frame. Candidate signal events are required to have a ``$1-3$ topology," meaning one $\tau$ decay yields three charged particles (3-prong), while the other $\tau$ decay yields one charged particle (1-prong). Selection of the $1-3$ prong decays begins by requiring events to have four well-reconstructed charged particles, none of which must be compatible with coming from a photon conversion track pair. The total charge of the four tracks must be zero.  Due to the large CM energy relative to the $\tau$ masses, the decay daughters of the two $\tau$s tend to be well-separated. Thus, the event is divided into two hemispheres in the CM frame by a plane perpendicular to the thrust axis, calculated using all observed charged and neutral particles in the event. One hemisphere is required to contain just one track and is termed the ``tag-side."  The other hemisphere (the ``signal-side") must include 3 charged tracks. All charged tracks are reconstructed assuming the pion mass hypothesis. In this analysis the 1-prong track must be identified as either leptonic channel, $\tau^{-} \rightarrow e^{-}\bar{\nu}_{e}\nu_{\tau}$ or  $\tau^{-} \rightarrow \mu^{-}\bar{\nu}_{\mu}\nu_{\tau}$. These two channels have a total branching fraction of $\sim 35\%$ and are chosen since they result in a better suppression of low-multiplicity $q\bar{q}$ background events. The terms ``electron tag" and ``muon tag"  refer to the leptons produced within the 1-prong channel. Each 1-prong channel is analyzed separately. The electrons are selected using a likelihood method and the muons are found using a set of selection criteria which employ information from all five sub-detectors.

\section{\label{sec:simulation}Simulation}

\subsection{Background Samples}

There are three source of potential backgrounds:

\begin{enumerate}
    \item  $\tau^{-} \rightarrow \pi^{-}\pi^{-}\pi^{+}\nu_{\tau}$, with an outgoing SM neutrino;
    \item other SM $\tau$ decays that have been misidentified as the 3-prong (3 charged pion) decay;
    \item non-$\tau$ backgrounds that have been misidentified as the 3-prong decay.
\end{enumerate}

In this analysis the SM background yields are estimated from Monte Carlo (MC) simulations, which are then passed through the same reconstruction and digitization routines as the data. The following sections describe the origins of each type of background and outline how MC simulations were generated. 

\subsubsection{\texorpdfstring{$\tau$}~ Backgrounds}

All $\tau$-pair events within \babar~ are simulated with higher-order radiative corrections using the KK2F MC generator \cite{kk2f}. The $\tau$-lepton decays are simulated using TAUOLA \cite{tauola}, which uses the averaged experimentally measured $\tau$ branching rates as listed in Ref. \cite{newpdg}.

Additional $\tau$ backgrounds originate when $\tau$-lepton decay modes other than the $\tau^{-} \rightarrow \pi^{+}\pi^{-}\pi^{-}\nu_{\tau}$ decay are misidentified as that channel. The largest contributions will come from channels with additional neutral particles: $\tau^{-} \rightarrow \pi^{+}\pi^{-}\pi^{-}\nu_{\tau} + N\pi^{0} $, where N = $1,2,3 ...$. The N $ =1$ channel will provide the largest contribution. A small amount of background events originate from kaon channels such as $\tau^{-} \rightarrow K^{+}K^{-}K^{-}  \nu_{\tau}$, $\tau^{-} \rightarrow 2K^{\pm}  \pi^{\pm} \nu_{\tau}$ and $\tau^{-} \rightarrow K^{-}  \pi^{-} \pi^{+} \nu_{\tau}$, where one or more charged kaons are tagged as pions. 

\subsubsection{non-\texorpdfstring{$\tau$}~ Backgrounds}

Several non-$\tau$ backgrounds are also possible, including:

\begin{itemize}
    \item  $e^{+}e^{-} \rightarrow \Upsilon(4S) \rightarrow B^{+} B^{-}$ and  $e^{+}e^{-} \rightarrow \Upsilon(4S) \rightarrow B^{0} \Bbar^{0}$, which are simulated using EvtGen \cite{EvtGen};
    \item  $e^{+}e^{-} \rightarrow  u\bar{u} ,  d\bar{d},s\bar{s}$ and   $e^{+}e^{-} \rightarrow  c\bar{c}$, which are simulated using JETSET \cite{ref2} \cite{ref3};
    \item $e^{+}e^{-} \rightarrow \mu^{+} \mu^{-} (\gamma)$, which are simulated using KK2F \cite{ref1}.
\end{itemize} 

Bhabha events, $e^{+}e^{-} \rightarrow e^{-} e^{+} (\gamma)$, are not a source of significant background after the event selection described below is applied. Any contamination from Bhabha events will be quantified and included as a systematic in the final analysis.

\subsection{Signal Samples}
\label{sec:signal_samples}
A total of 26 signal samples were simulated, one for each of the HNL masses across the range 100 \mevcc $< m_{4} <$ 1300 \mevcc, at 100 \mevcc~ increments. For each of these HNL masses both a $\tau^{+}$ and $\tau^{-}$ signal channel were simulated. We assume a HNL could appear in either channel. 

Signal samples were produced within the \babar~ software environment using KK2F and TAUOLA by changing the value of the outgoing neutrino mass in TAUOLA.  The generated signal was then passed through the same digitization and reconstruction model as the SM background and data samples. 
 
\begin{figure}[t]
    \centering
    \includegraphics[width=3in,height=2.0in]{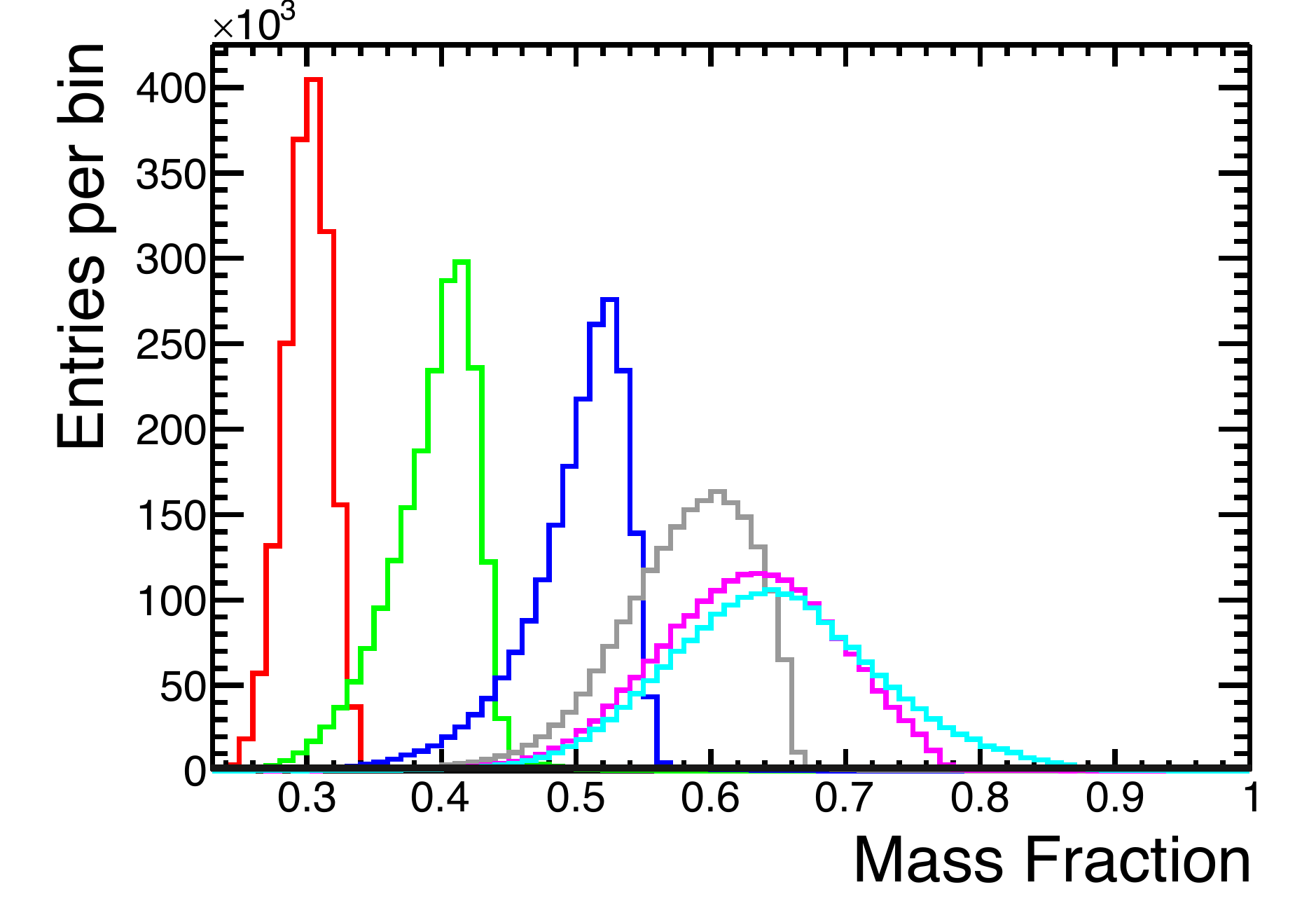}
    \includegraphics[width=3in]{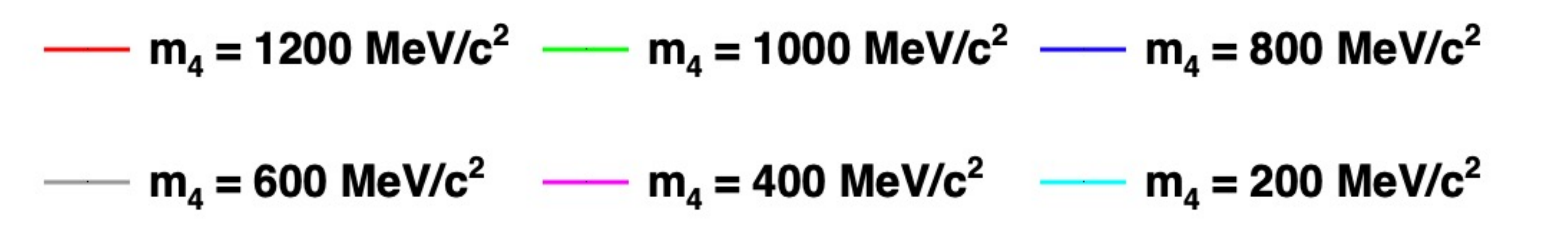}
     \includegraphics[width=3in,height=2.0in]{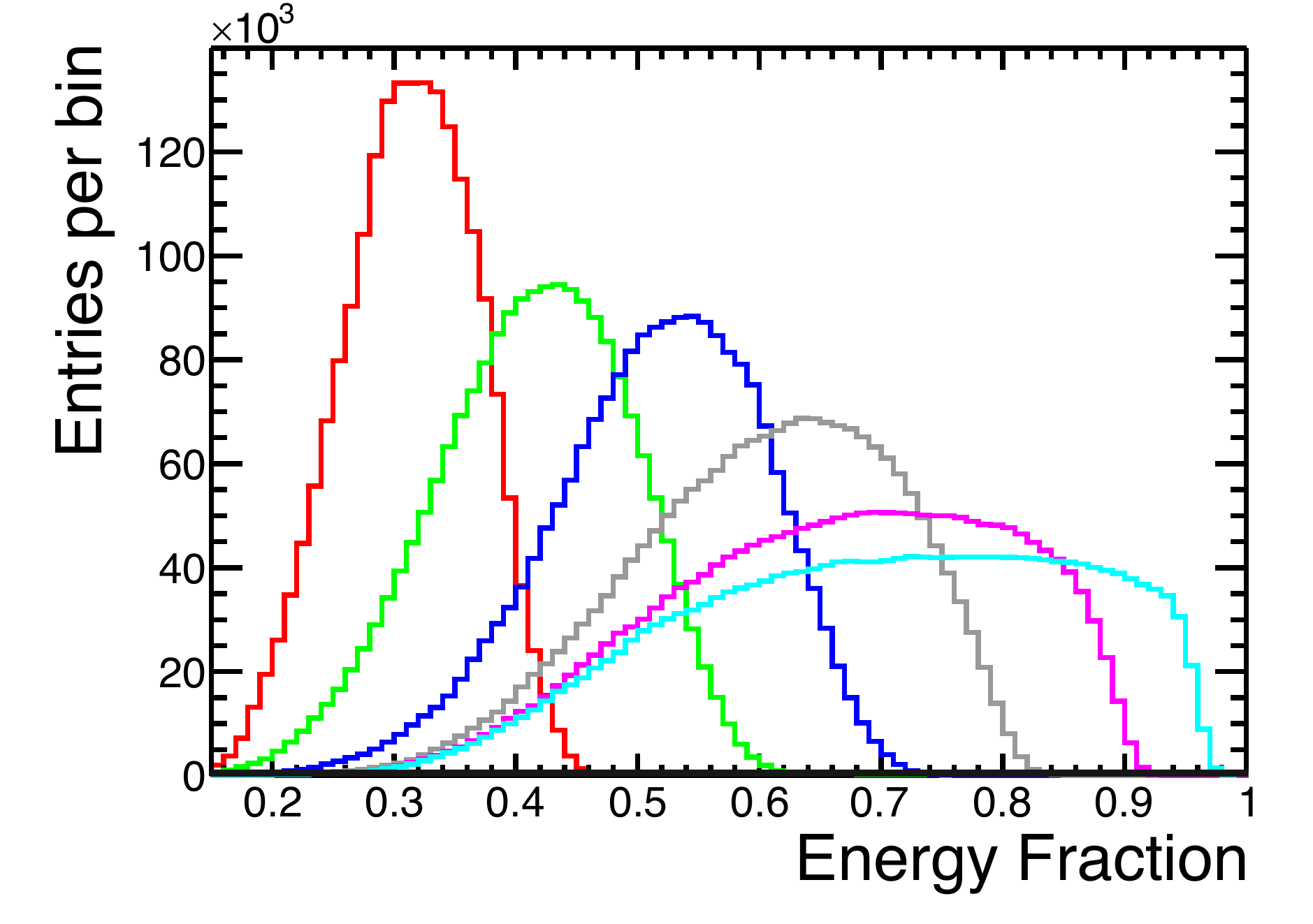}
      \includegraphics[width=3in]{figs/Legend.pdf}
    \caption{1D projections of the (top) reconstructed invariant-mass and (bottom) reconstructed energy of the outgoing hadronic system from $\tau^{-} \rightarrow \pi^{-} \pi^{-} \pi^{+} + \nu_{4} $, as fractions of that of the $\tau$-lepton, for some of the HNL mass hypotheses studied. Samples are normalized for a nominal $2\times 10^{6}$ reconstructed signal events to allow direct comparisons between the samples.}
    \label{fig:sig_EM}
\end{figure}

\begin{figure*}[t]
\centering
\begin{tabular}{cc}
 $m_{4}$ = 0 \mevcc & $m_{4}$ =500 \mevcc \\
\includegraphics[width=3in,height=2in]{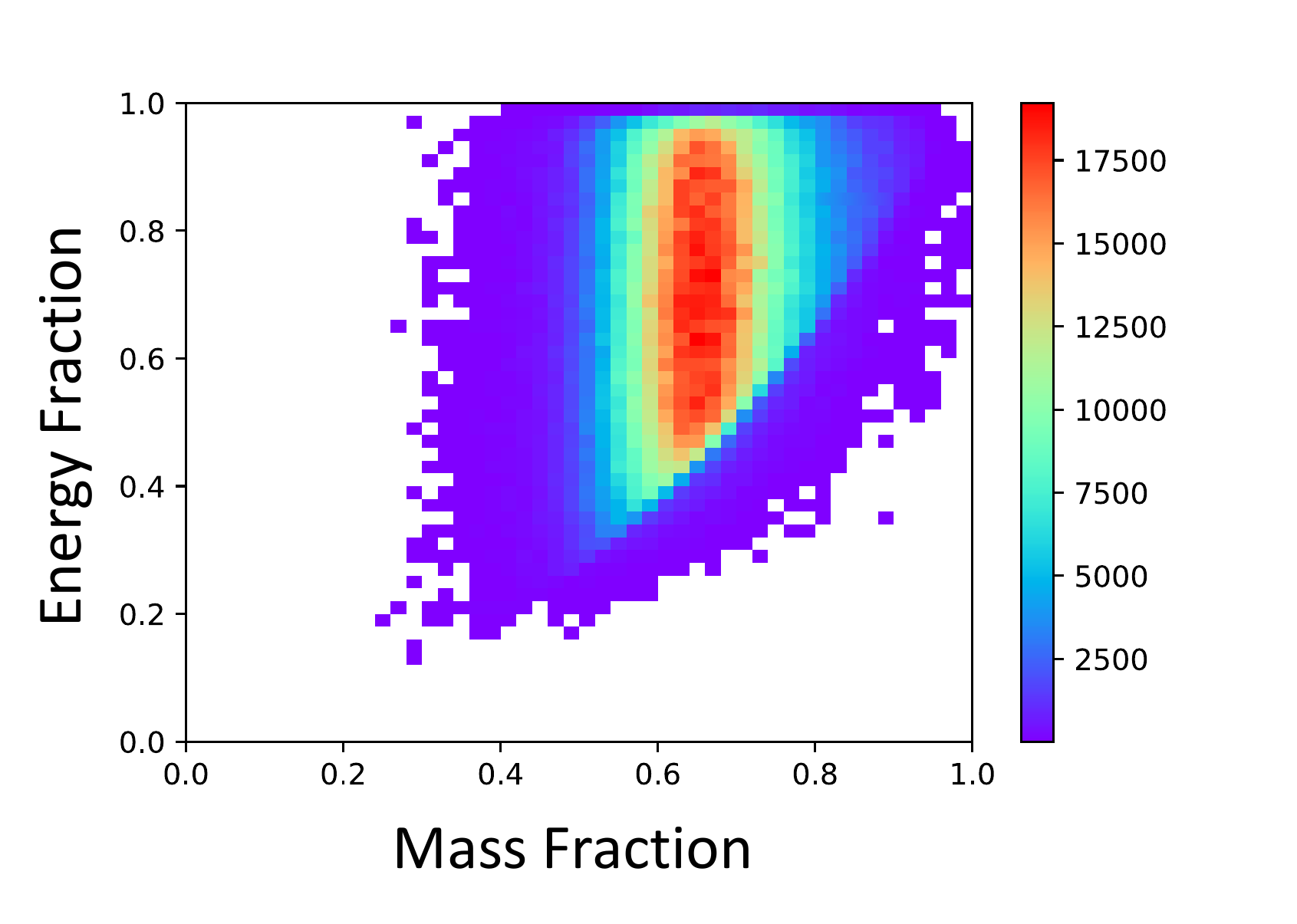}
& \includegraphics[width=3in,height=2in]{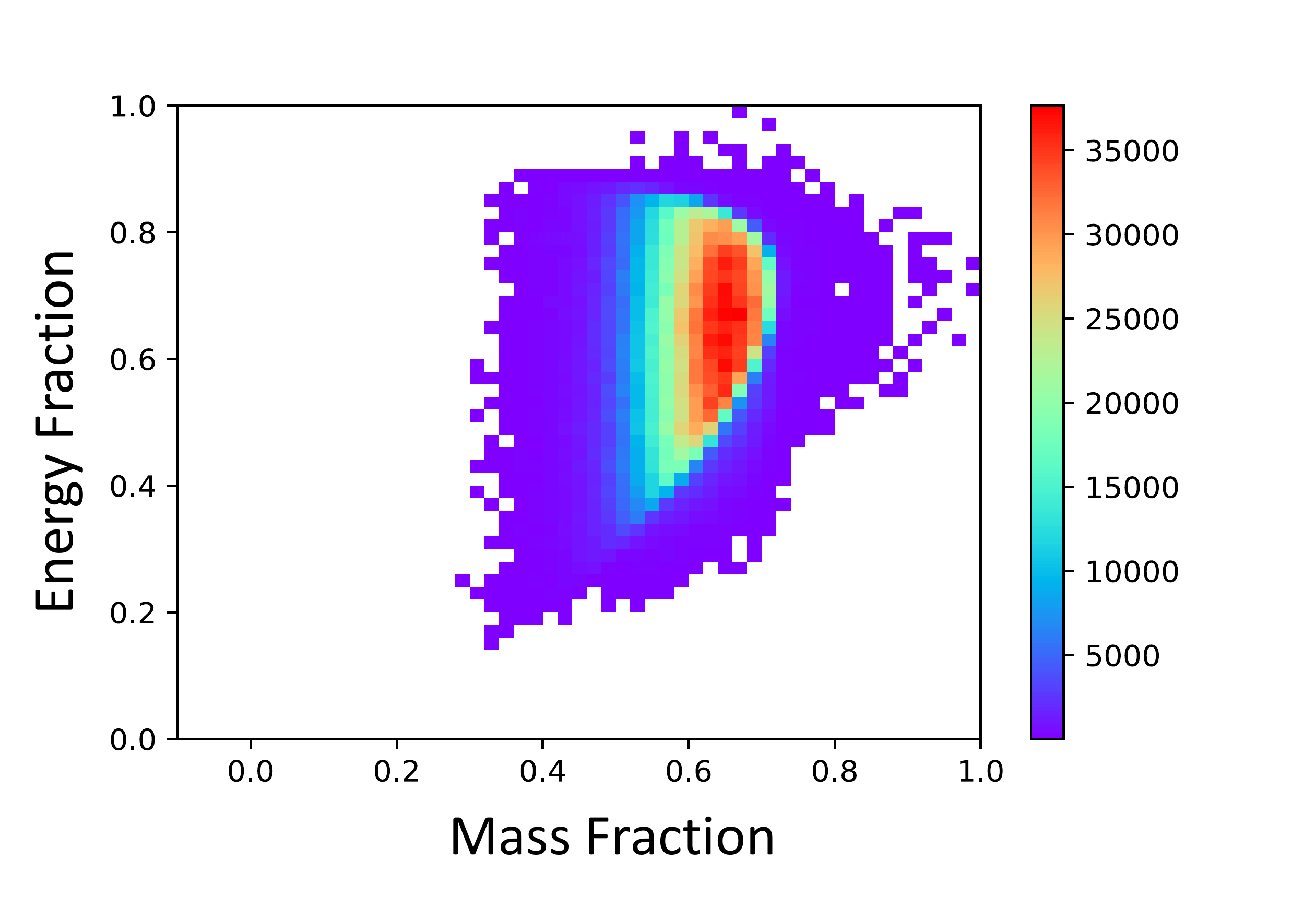}\\
  $m_{4}$ =700 \mevcc &  $m_{4}$ =1000 \mevcc \\
\includegraphics[width=3in,height=2in]{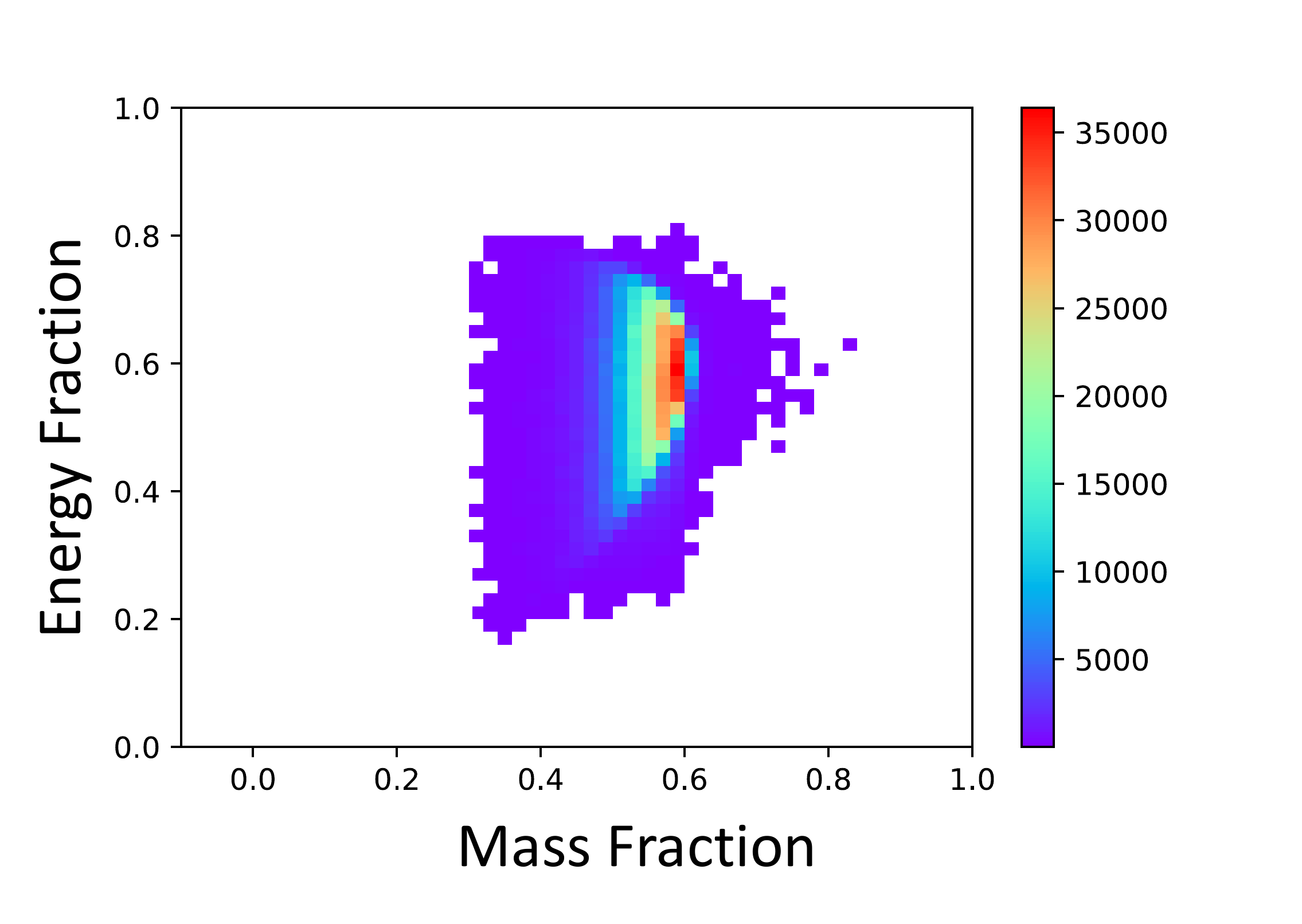}
 & \includegraphics[width=3in,height=2in]{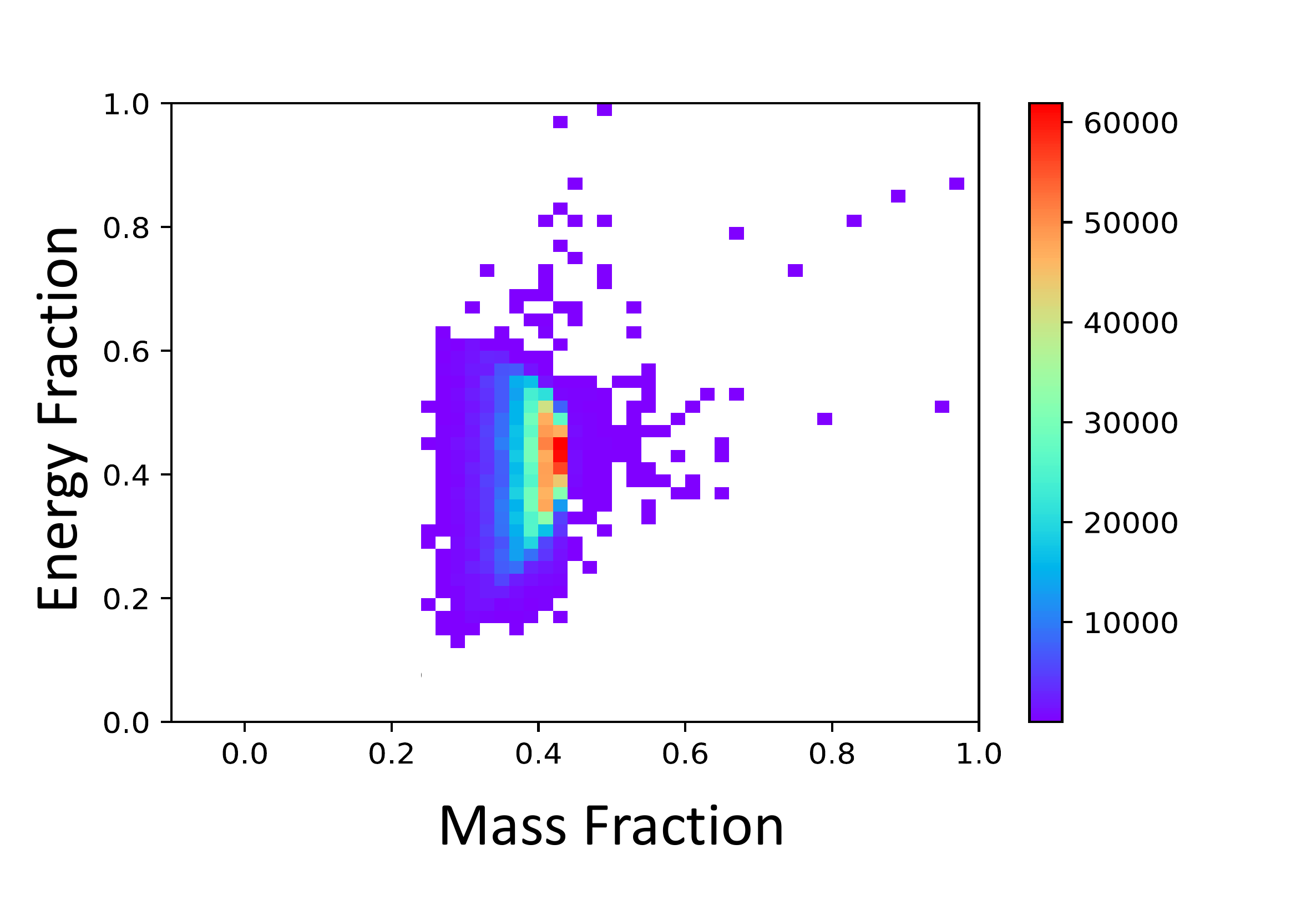}\\
\end{tabular}
\caption{Example of reconstructed invariant-mass and energy ($m_{h},E_{h}$) (as fraction of incoming $\tau$ mass and energy) of the outgoing hadronic system from $\tau^{-} \rightarrow \pi^{-} \pi^{-} \pi^{+} + \nu_{4} $ for HNL  masses of 100, 500, 700 and 1000 \mevcc. No selection criteria are applied other than particle identification. The simulated signal samples in these plots assume a non-signal (tag) $\tau$-lepton decaying leptonically to an electron and SM neutrinos.}
        \label{fig:sig_templates}
\end{figure*}

Figure~\ref{fig:sig_EM} shows a few 1D projections of the reconstructed invariant-mass ($m_{h}$) and CM energy ($E_{h}$) of the outgoing hadronic system, as fractions of that of the $\tau$-lepton, for various HNL mass hypotheses. The signal samples all have $2 \times 10^{6}$ reconstructed events to allow direct comparisons of the differences in shape. The reason for acquiring such high statistics is to help reduce statistical fluctuations and to ensure that tail regions are well populated. As the HNL mass increases the fraction of the tau's energy and mass going to the visible, hadronic, system decreases. The same effect is visible in Fig.~\ref{fig:sig_templates} which shows examples of the 2D reconstructed templates.

\subsection{Digitization modeling, Triggering and Reconstruction}

 The propagation of particles through the detector geometries is simulated using the GEANT4 toolkit \cite{Geant}. \babar-specific software is then used to convert the GEANT4 outcome into simulated digitized raw data, which was then subjected to full reconstruction. In this analysis simulated events for HNL signal and SM processes are reconstructed in the same manner as the data. The simulation takes into account the variation of the detector and trigger configurations as well as the accelerator conditions and the beam-induced background. Trigger and filter algorithms corresponding to those applied to the real data are applied to all the simulated data. Real-data selected events, corresponding to non-colliding-beam backgrounds from the accelerator, are mixed with simulated events.

\section{\label{sec:selection}Event Selection}

Section~\ref{exp_strat} details the initial event selection and how the topology of the events are selected. In addition, backgrounds from $q\bar{q}$ and two-photon events have a more spherically symmetric shape and lower thrust than the events from $\tau$-pairs. Therefore, a thrust higher than 0.85 is required. This reduces contamination from $e^{+}e^{-} \rightarrow q\bar{q}$ backgrounds to $\sim$ 0.1$\%$ of the 3-prong candidates. This thrust requirement removes only $\sim$ 4$\%$ of all $\tau$ events. To reject events that do not have missing particles the $p_{T}$ is required to be greater than $0.9\%$ of the CM energy. A requirement on the missing momentum in the CM frame of $> 0.9$\%$ \sqrt{s}$ is enforced to suppress the remaining non-$\tau$ background events, including those from two-photon processes. In order to ensure a good particle identification (PID) performance, each track is required to be within the geometrical acceptance of the DIRC and the EMC: $-0.76 < \cos(\theta) < 0.9$. In addition, the track must have a minimum $p_{T}$ of 250 MeV/$c$, enabling them to reach the DIRC. The invariant-mass of the 3-prong system, $m_{123}$, must be less than or equal to the $\tau$-lepton mass ($1.776$ \gevcc). 

Any event containing a track compatible with being a daughter of a converted photon  \cite{Lurie_2002} is rejected: if $m_{\gamma} < 15$ \mevcc, where $m_{\gamma}$ is the converted photon candidate mass, and $\delta_{xy} < 0.5$ cm, $\delta_{z} < 1.0$ cm, and $r_{xy}/\sigma_{r_{xy}} > 2.5$; where $\delta_{xy}$ and $\delta_{z}$ are the distances of closest approach of the charged tracks from the conversion in the transverse and longitudinal directions, respectively, and $r_{xy}/\sigma_{r_{xy}}$ is the ratio of the fitted vertex decay length to the error on that value. 

Events containing a $K^{0}_{S}$ are considered to be backgrounds. Candidate $K^{0}_{S}$ events are identified as a pair of oppositely charged particles in the signal hemisphere with a di-pion invariant-mass consistent with $K^{0}_{S}$ mass, and with a decay length in the $xy$-plane greater than six standard deviations. These candidates are rejected. In order to reduce backgrounds from events containing neutral particles we apply specific selection criteria to discriminate against reconstructed energy deposits in the EMC not associated with a charged track. For the leptonic, 1-prong side neutral clusters are associated with the track if:

\begin{itemize}
    \item \textbf{For Electrons}  the neutral cluster has $E^{\text{neut,1-prong}}_{\text{EMC}}$ $< 1 - 0.016 \frac{\text{GeV}}{\text{cm}}d$ where $d < 50$cm  or, if $d > 50$ cm, $E^{\text{neut}}_{\text{EMC}} < 0.2$ GeV, 
    \item \textbf{For Muons} the neutral cluster has $E^{\text{neut,1-prong}}_{\text{EMC}} < 0.2$ GeV,
\end{itemize}
where $E^{\text{neut,1-prong}}_{\text{EMC}}$ is the measured EMC energy of the extra cluster and $d$ is the distance between the  EMC cluster associated to the track and the nearest neutral cluster. The remaining neutral particles are considered to be un-associated with the cluster and the event is vetoed if:

\begin{itemize}
    \item \textbf{For Electrons} in the 1-prong case there is an un-associated neutral cluster with $E^{\text{neut, 1-prong}}_{\text{EMC}} > 1$ GeV or if the 3-prong side has $\sum E^{\text{neut,3-prong}}_{\text{EMC}} > 0.2$ GeV.
    \item \textbf{For Muons} in the 1-prong side there is an un-associated neutral cluster with $E^{\text{neut, 1-prong}}_{\text{EMC}}$ $> 0.5$ GeV or if the 3-prong side has $\sum E^{\text{neut,3-prong}}_{\text{EMC}} > 0.2$ GeV.
\end{itemize}

\section{\label{sec:yield}Expected Yield}

Tables~\ref{table:bkg_pos} and \ref{table:bkg_neg} list the non-negligible background yields, calculated using reconstructed MC samples (as described in Sec.~\ref{sec:simulation}), with all selection requirements applied. By far the largest source of background comes from the SM 3-prong $\tau$ decay, that provides $\sim$ 70$\%$ of the total events. The 3-prong SM hadronic decay, accompanied by 1 or 2 neutral pions provides $\sim 27\%$ of the events, the second largest contribution. The channel with two charged pions and a charged kaon, where the kaon is tagged as a pion, makes up the fourth highest contribution of the $\tau$ backgrounds. Yields from $B^{0}\Bbar^{0}$, $B^{+}B^{-}$ and $\mu^{-}\mu^{+}$  are expected to be very low, with the latter being completely negligible in the electron channel. The $q\bar{q}$ backgrounds make up a total $\sim 1-2\%$ of the overall expected background yield.  

The MC simulation yields exceed those from data by $0.48$ to  $0.99$$\%$.

\begin{table*}[t!]
\centering
\caption{List of expected background yields after all selection requirements are applied for positive 3-prong channel. All backgrounds are scaled to represent what would be expected for $\mathcal{L} = 424 \text{fb}^{-1}$.}
\begin{tabular}{l |ll| ll} 
& Electron Tag && Muon Tag &\\
Bkg. Type  & MC Yield  &  [$\%$]& MC Yield  &[$\%$] \\ [0.5ex] 
 \hline \hline
 $\tau^{+} \rightarrow \pi^{+}\pi^{+}\pi^{-}\bar{\nu}_{\tau}$&894864&70&810586&71\\
 \hline
 $\tau^{+} \rightarrow \pi^{+}\pi^{+}\pi^{-}\pi^{0}  \bar{\nu}_{\tau}$&332008&26&278830&24\\
 $\tau^{+} \rightarrow \pi^{+}\pi^{+}\pi^{-} 2\pi^{0}  \bar{\nu}_{\tau}$&34050&2.7&28841&2.5\\
   $\tau^{+} \rightarrow \pi^{-}\pi^{+}  K^{+} \bar{\nu}_{\tau}$ &3391&0.27&3101&0.27\\
 $\tau^{+} \rightarrow \pi^{+}\pi^{+}\pi^{-}  3\pi^{0}  \bar{\nu}_{\tau}$&1541&0.12&821& 0.07\\
  $\tau^{+} \rightarrow \pi^{+}  \pi^{0}  \bar{\nu}_{\tau}$ &498&0.039&207&0.017\\
 $\tau^{+} \rightarrow \pi^{+}  2\pi^{0}  \bar{\nu}_{\tau}$ &252&0.02&92&0.27\\
 $\tau^{+} \rightarrow 2K^{+}  \pi^{-}  \bar{\nu}_{\tau}(\rightarrow K^{+} K^{-} \pi^{+}  \bar{\nu}_{\tau})$&207&0.016&146& 0.013\\

 \hline
 $e^{+}e^{-} \rightarrow Y(4S) \rightarrow c\bar{c}$&8031&0.63&6512&0.55\\
 $e^{+}e^{-} \rightarrow Y(4S) \rightarrow u\bar{u} , s\bar{s}, d\bar{d}$&542&0.043&13898&1.19\\
 $e^{+}e^{-} \rightarrow Y(4S) \rightarrow B^{0} \Bbar^{0}$&108&0.009&99&0.0084 \\
$e^{+}e^{-} \rightarrow Y(4S) \rightarrow B^{+} B^{-}$ &100&0.008& 89&0.0076\\
$e^{+}e^{-} \rightarrow \mu^{+}\mu^{-}$ &0&0&15&0.0013\\
\hline
\hline
\textbf{Total MC}&1278339&-&1143237&-\\
\textbf{Data} &1265698&-&1137521&-\\
   \hline
\end{tabular}
\label{table:bkg_pos}
\end{table*}

\begin{table*}[t!]
\centering
\caption{List of expected background yields after all selection requirements are applied for negative 3-prong channel. All backgrounds are scaled to represent what would be expected for $\mathcal{L} = 424 \text{fb}^{-1}$.}
\begin{tabular}{l |ll |ll} 
& Electron Tag && Muon Tag&\\
 Bkg. Type  & MC Yield &  [$\%$]& MC Yield  &[$\%$] \\ [0.5ex] 
 \hline \hline
 $\tau^{-} \rightarrow \pi^{-}\pi^{-}\pi^{+} \nu_{\tau}$&900069&70&817342&70\\
 \hline
 $\tau^{-} \rightarrow \pi^{-}\pi^{-}\pi^{+}  \pi^{0}  \nu_{\tau}$&334565&26&281613&25\\
 $\tau^{-} \rightarrow \pi^{-}\pi^{-}\pi^{+}  2\pi^{0}  \nu_{\tau}$&34255&2.7&29287&2.5\\
   $\tau^{-} \rightarrow \pi^{+} \pi^{-}  K^{-}  \nu_{\tau}$ &3567&0.27&3228&0.27\\
 $\tau^{-} \rightarrow \pi^{-}\pi^{-}\pi^{+}  3\pi^{0}  \nu_{\tau}$&1535&0.12&795& 0.07\\
  $\tau^{-} \rightarrow \pi^{-}  \pi^{0}  \nu_{\tau}$ &476&0.039&217&0.019\\
 $\tau^{-} \rightarrow \pi^{-}  2\pi^{0} \nu_{\tau}$ &240&0.02&92&0.08\\
$\tau^{-} \rightarrow 2K^{-}  \pi^{+}  \bar{\nu}_{\tau}(\rightarrow K^{-} K^{+} \pi^{-}  \bar{\nu}_{\tau})$&202&0.016&152& 0.013\\

 \hline
 $e^{+}e^{-} \rightarrow Y(4S) \rightarrow c\bar{c}$&8031&0.63&6837&0.58\\
 $e^{+}e^{-} \rightarrow Y(4S) \rightarrow u\bar{u} , s\bar{s}, d\bar{d}$&495&0.043&16602&1.42\\
 $e^{+}e^{-} \rightarrow Y(4S) \rightarrow B^{0} \Bbar^{0}$&126&0.009&98&0.0083 \\
$e^{+}e^{-} \rightarrow Y(4S) \rightarrow B^{+} B^{-}$ &93&0.008& 103&0.0088\\
$e^{+}e^{-} \rightarrow \mu^{+}\mu^{-}$ &0&0&10&0.0009\\
\hline
\hline
\textbf{Total MC}&1283654&-&1155920&-\\
\textbf{Data} &1273291&-&1150350&-\\
   \hline
\end{tabular}
\label{table:bkg_neg}
\end{table*}

\subsection{Energy and Mass Distributions}

\begin{figure*}[t]
  \centering
  \includegraphics[width=3.5in,height=2.6in]{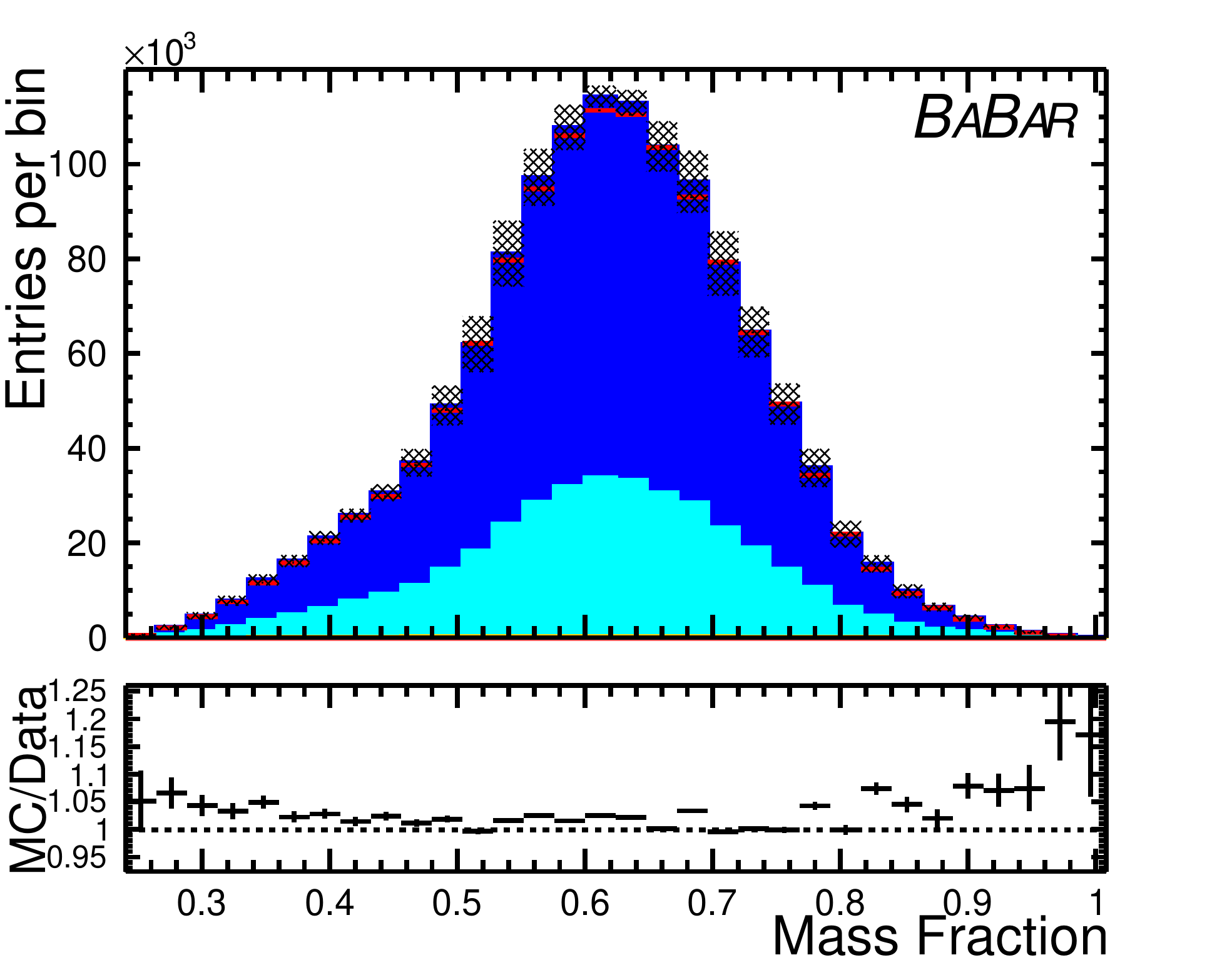}
    \includegraphics[width=3.5in,height=2.6in]{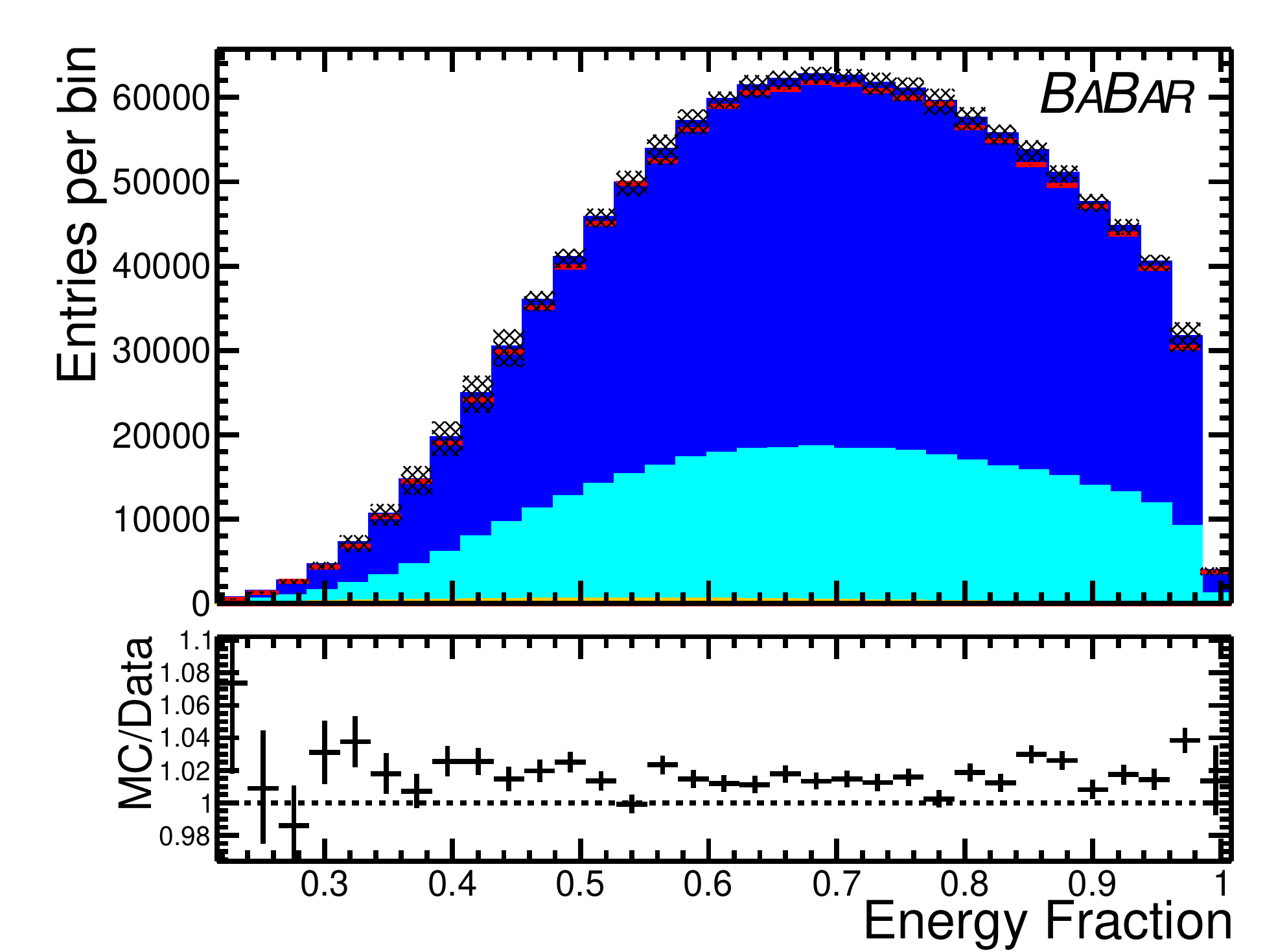}
      \includegraphics[width=3.5in,height=2.6in]{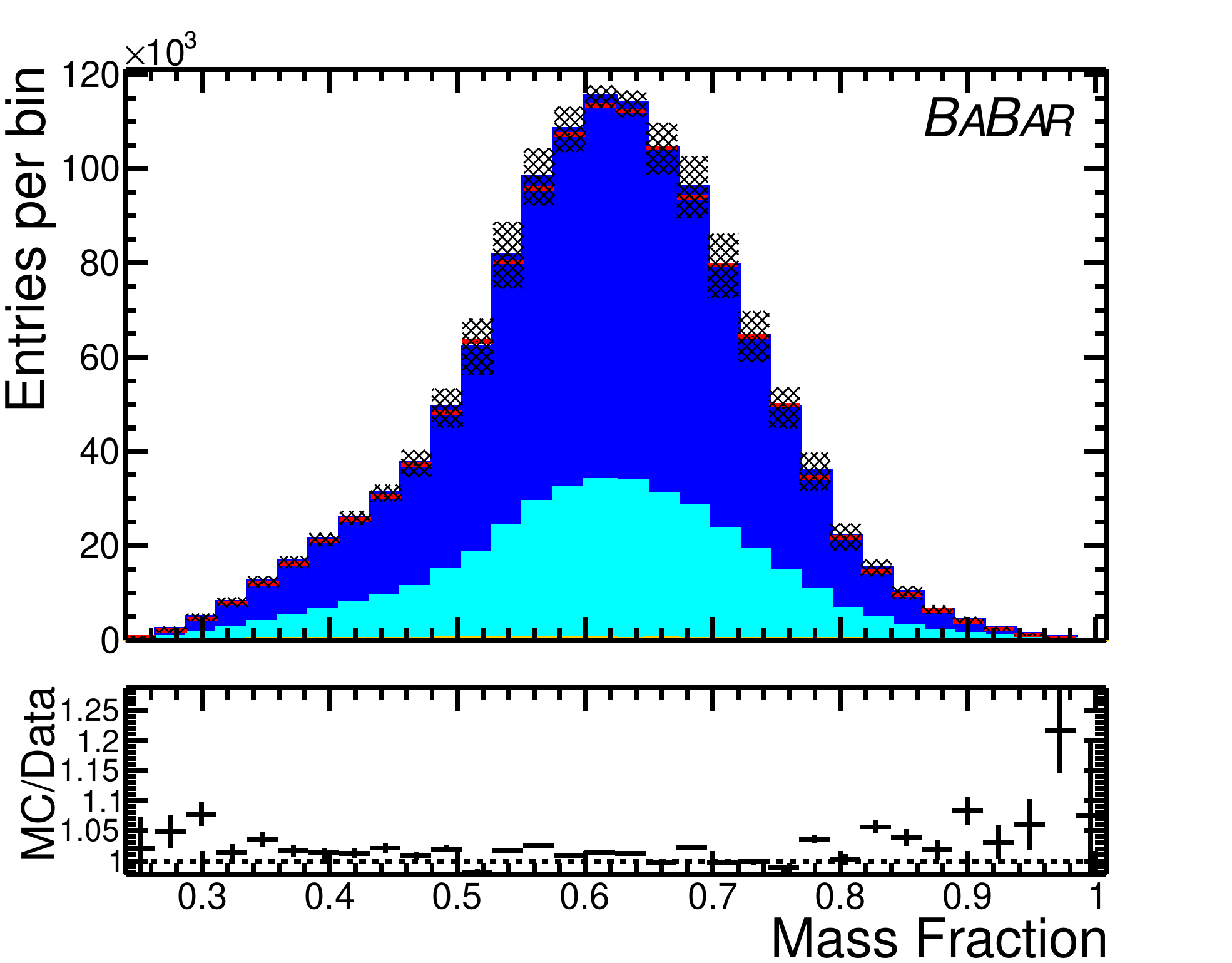}
    \includegraphics[width=3.5in,height=2.6in]{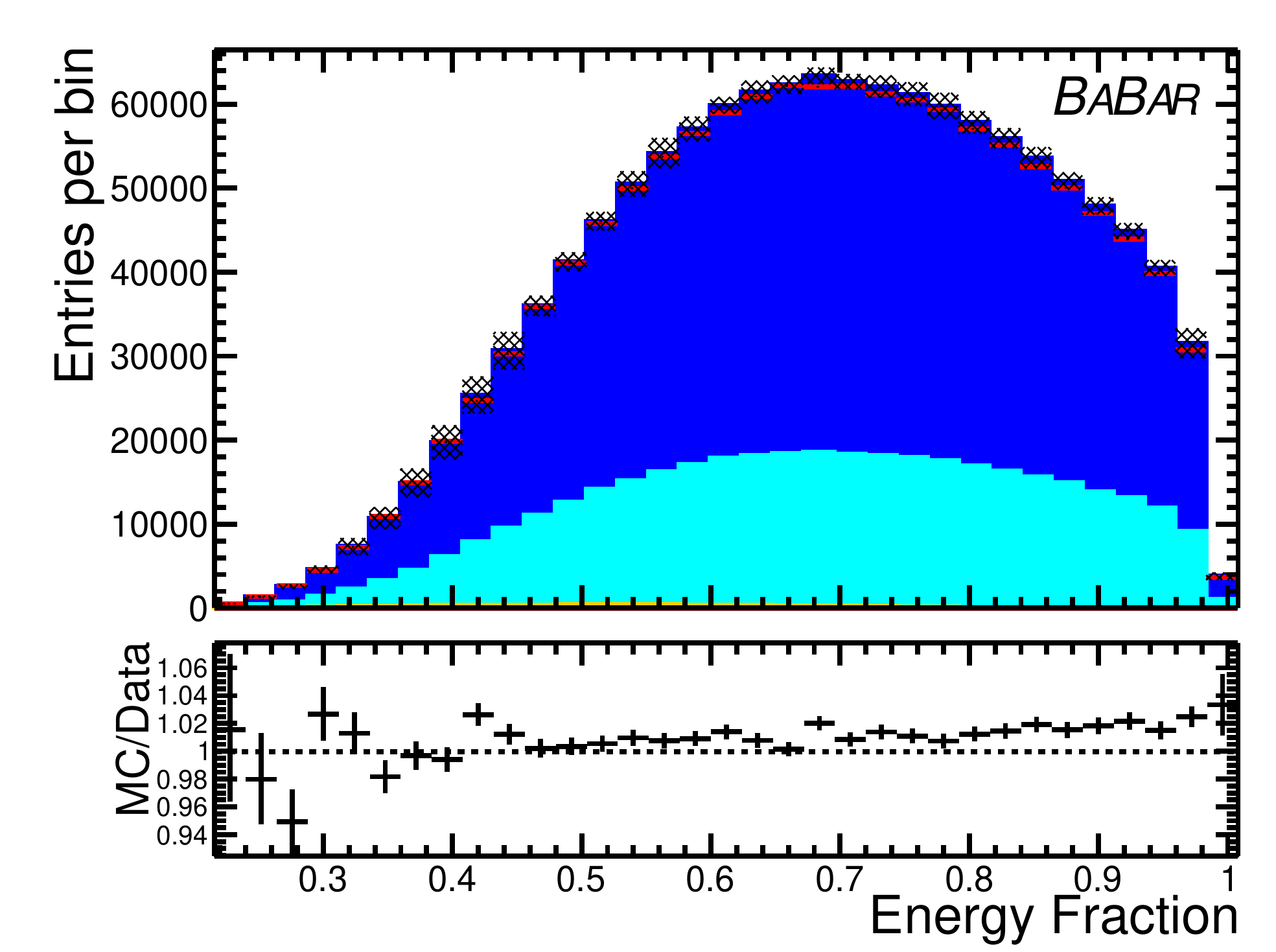}
    \includegraphics[width = 5.2in,height=0.6in]{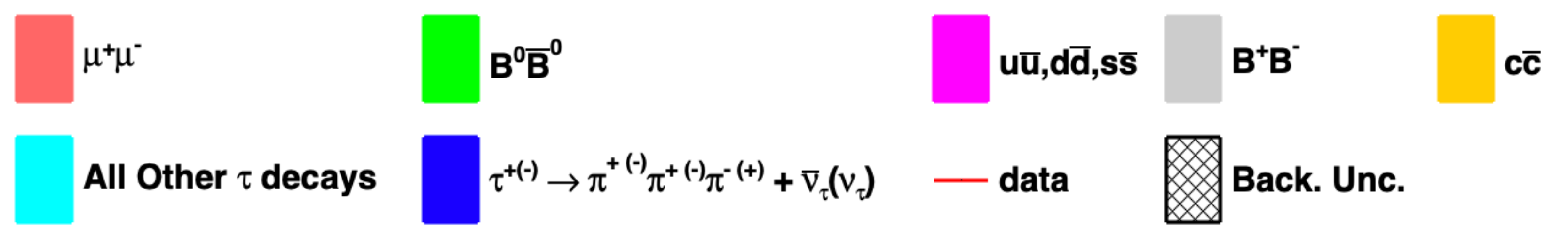}
  \caption{
  \label{fig:e_both}
  \textbf{Electron Tag}:  Reconstructed mass (left) and energy (right) of the outgoing hadronic system, as fractions of that of the incoming $\tau$-lepton, for the (top) positive $\tau$ signal channel and (bottom)  negative $\tau$ signal channel. The ``All other $\tau$" component contains all other $\tau$ decays which are not $\tau^{+(-)} \rightarrow \pi^{-(+)}\pi^{+(-)}\pi^{+(-)}+\bar{\nu}_{\tau}(\nu_{\tau})$. The individual contributions are detailed in Tables~\ref{table:bkg_pos}-~\ref{table:bkg_neg}. The lower figures show ratios of the total MC yield to data yield, in each bin. The error bars are statistical.}
  \end{figure*}

\begin{figure*}[t]
  \centering
  \includegraphics[width=3.5in,height=2.6in]{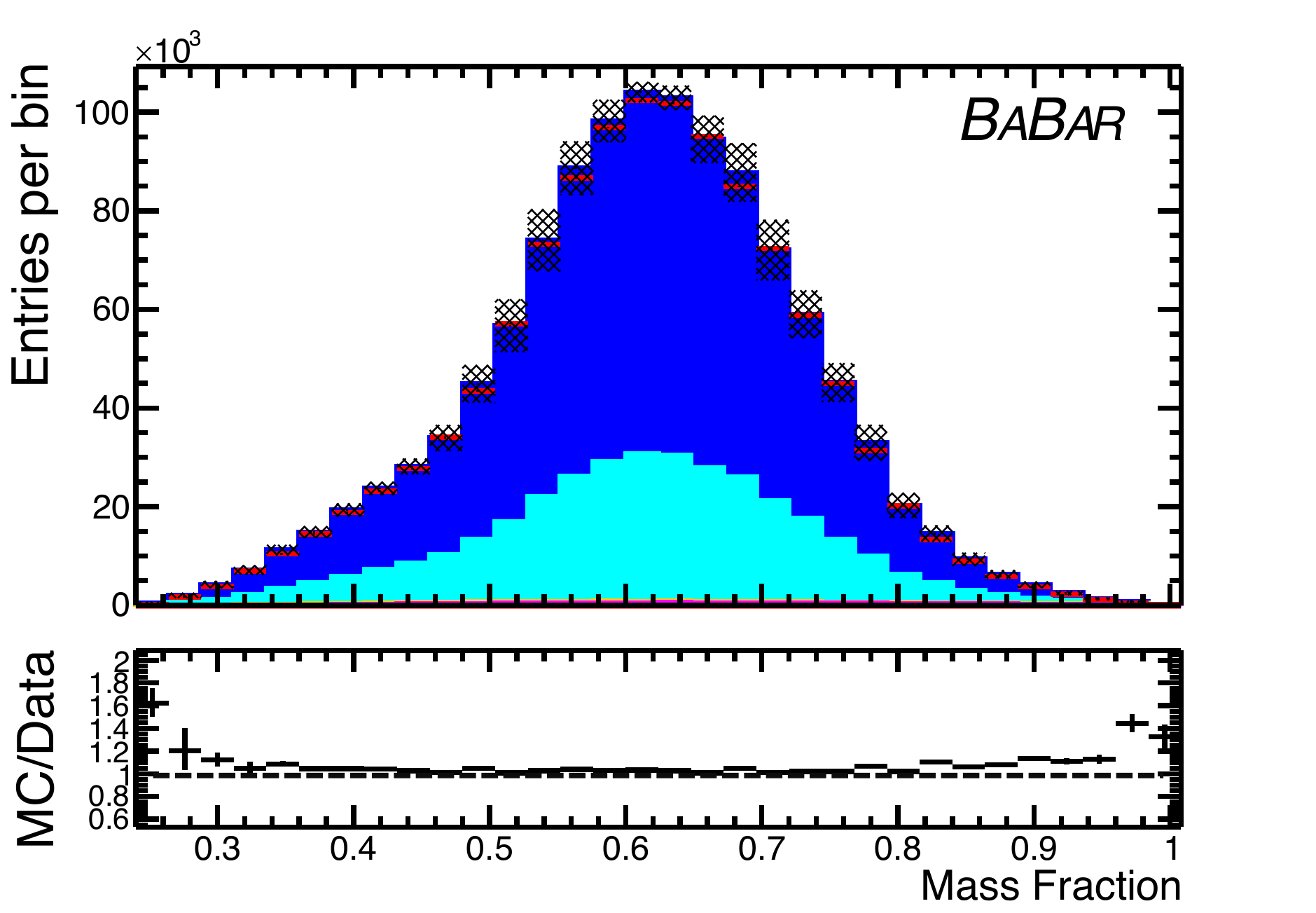}
    \includegraphics[width=3.5in,height=2.6in]{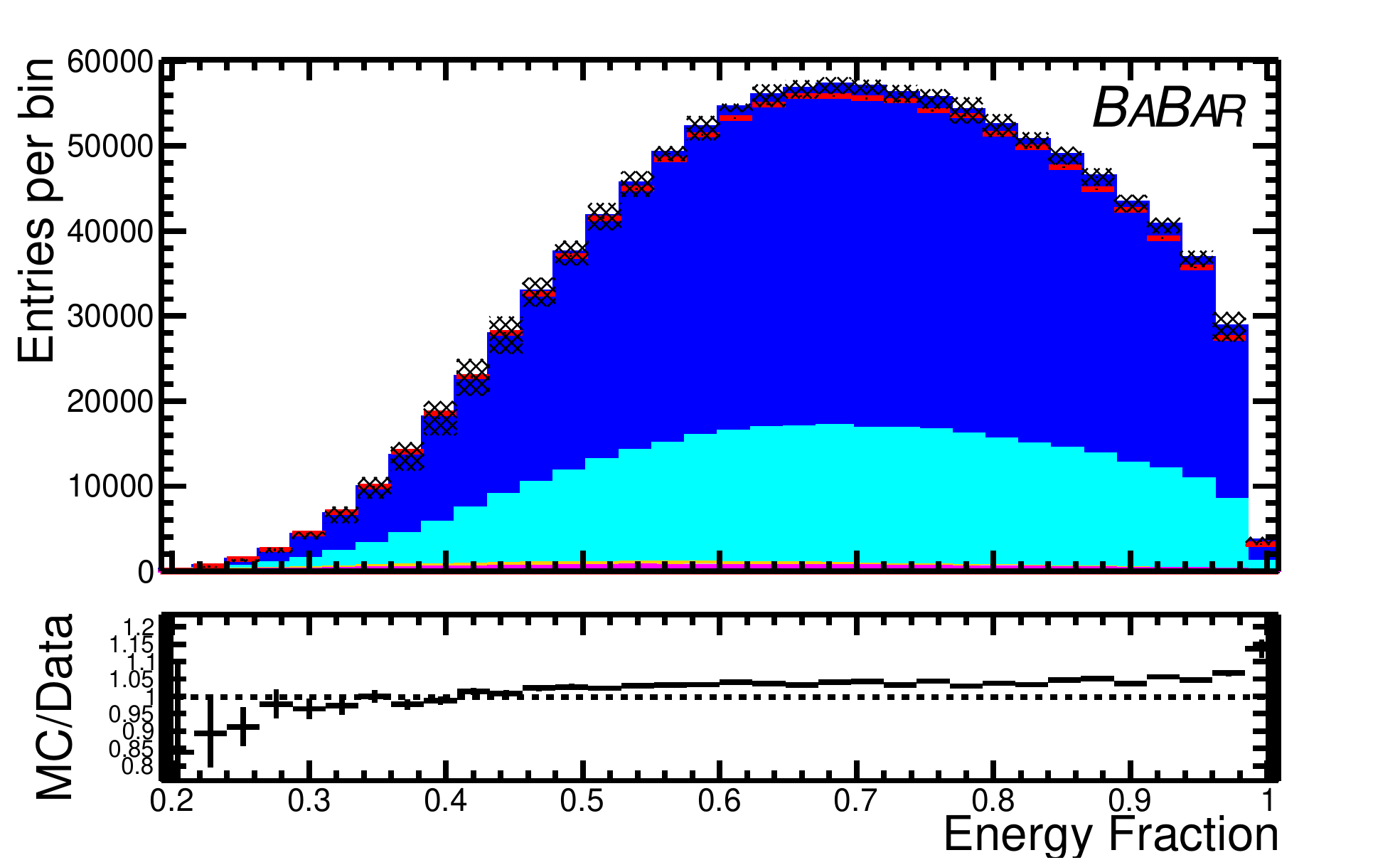}
      \includegraphics[width=3.5in,height=2.6in]{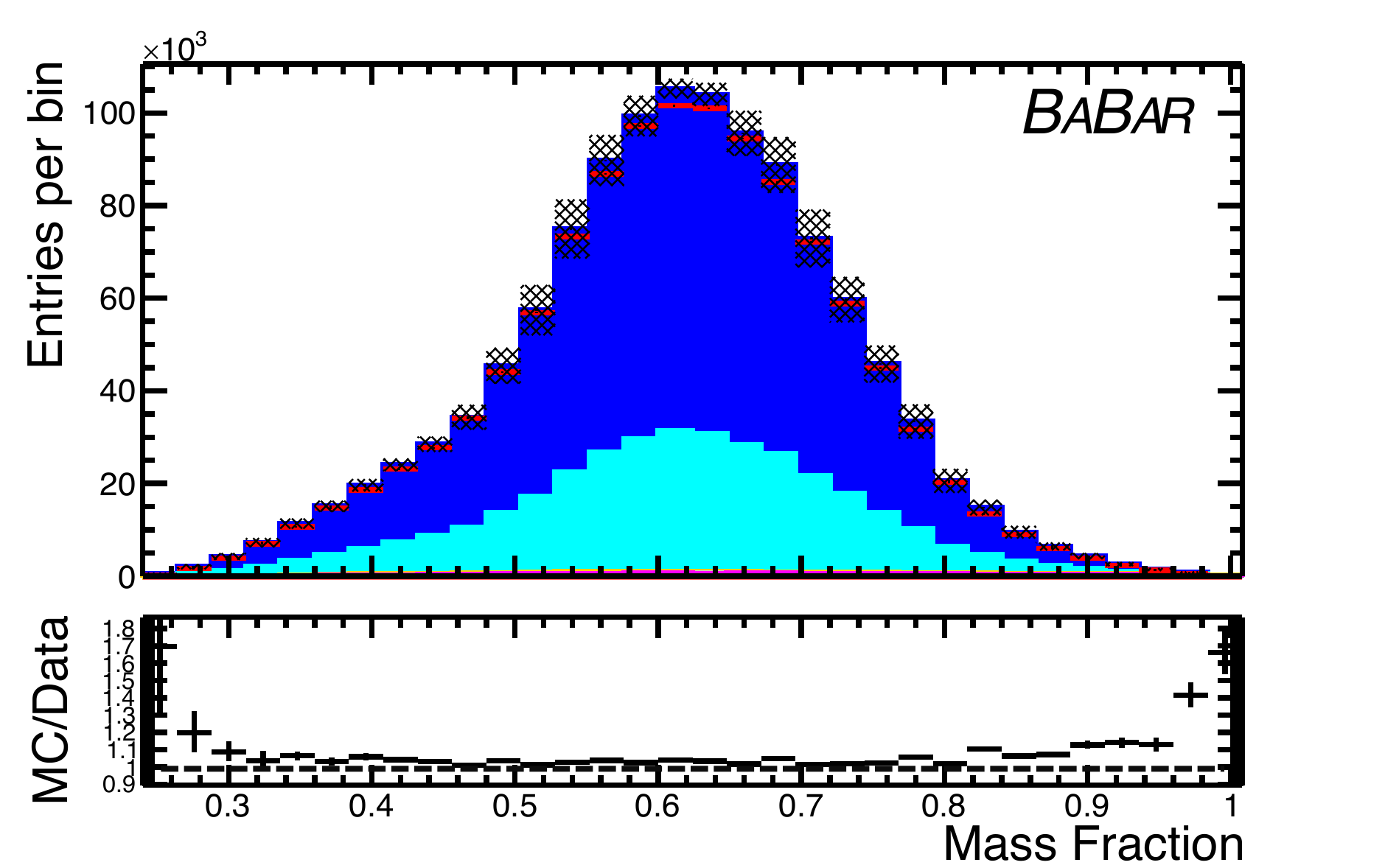}
    \includegraphics[width=3.5in,height=2.6in]{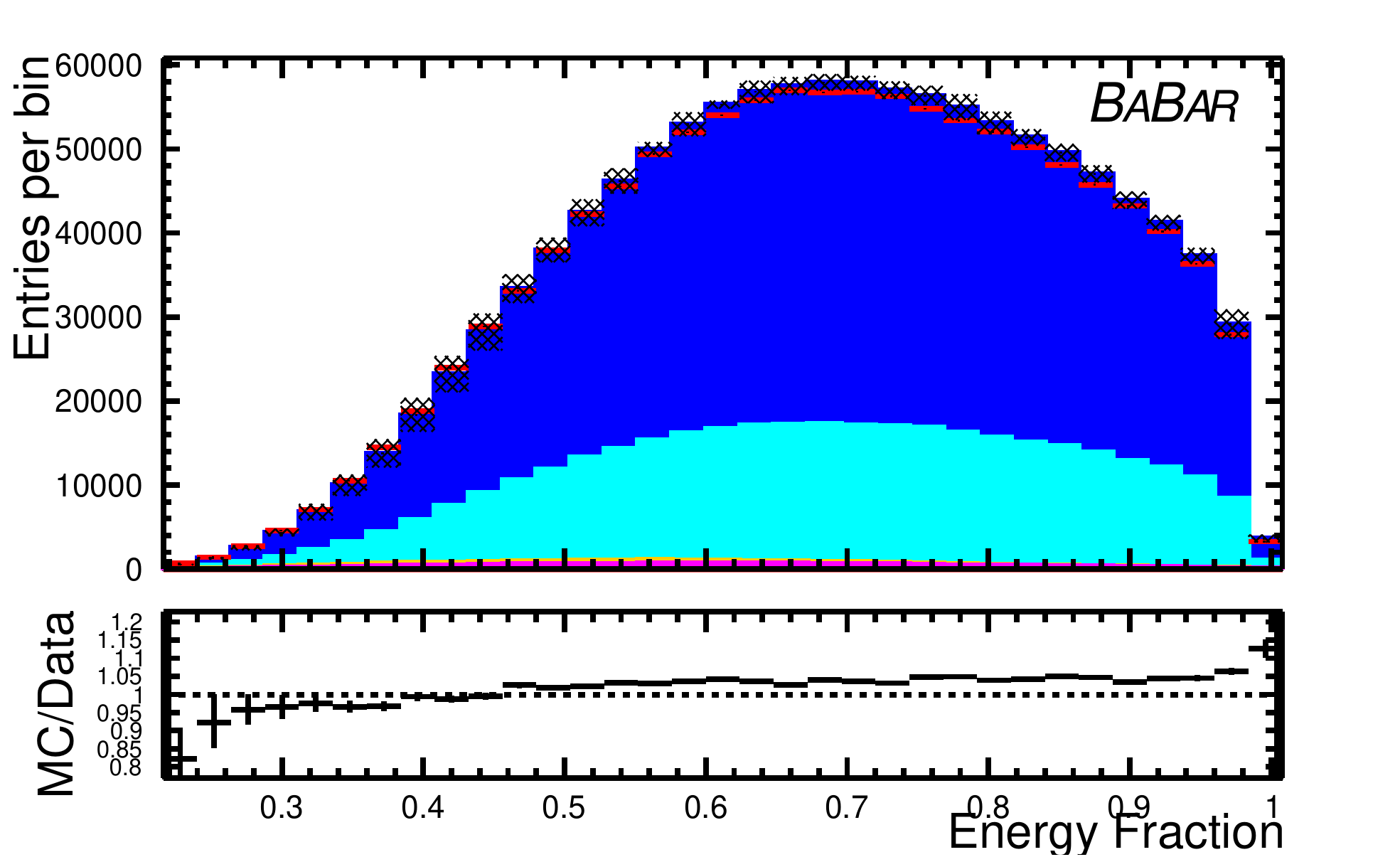}
    \includegraphics[width = 5.2in,height=0.6in]{figs/1DLegend.pdf}
  \caption{
  \label{fig:mu_both}
  \textbf{Muon Tag}: Reconstructed mass (left) and energy (right) of the outgoing hadronic system, as fractions of that of the incoming $\tau$-lepton, for the (top) positive $\tau$ signal channel and (bottom) negative $\tau$ signal channel. The ``All other $\tau$" component contains all other $\tau$ decays which are not $\tau^{+(-)} \rightarrow \pi^{-(+)}\pi^{+(-)}\pi^{+(-)}+\bar{\nu}_{\tau}(\nu_{\tau})$. The individual contributions are detailed in Tables ~\ref{table:bkg_pos}-~\ref{table:bkg_neg}.}
  \end{figure*}

Figures ~\ref{fig:e_both} and~\ref{fig:mu_both} show the reconstructed invariant-mass and energy fraction distributions of the hadronic system for events with an electron and muon 1-prong tag, respectively. The background uncertainty corresponds to the total uncertainty ($\pm 1 \sigma_{\text{total}}$) on each bin, which includes statistical uncertainty, uncertainty on $\tau$ branching fractions and uncertainties on the modeling of the underlying resonances. To account for the latter, the $\tau$ background MC distributions are re-weighted to reflect an underlying resonance mass and width values at either $\pm 1 \sigma$, where $\sigma$ is the averaged experimental uncertainty. This will be discussed in detail in Secs.~\ref{a1} -~\ref{sec:tau_shape}. The results of this analysis will quote two sets of limits, one which does not take into account these modeling uncertainties (and uses the average values presented in Ref. \cite{newpdg}), and a more conservative limit, which takes into account the largest possible deviation in the template histograms when the uncertainty on the modeling is accounted for. The effect of the change in the characteristics of the intermediate resonance is that events can shift into neighbouring bins. This could provide a dependence on bin size (see Sec.~\ref{sec:syserrors}).

The invariant-mass and energy distributions of the negative and positive signal channels, for the same lepton tag, are statistically consistent. The ratio plots included in these figures quantify the ratio of MC to data yield in each bin, along with statistical uncertainty on that ratio. When the background uncertainty is taken into account the MC and data distributions can be considered consistent in most bins. The mass and energy distributions associated with the electron tag and the muon tag are also similar in shape, which is to be expected. Slight differences appear due to differences in the PID algorithms and background content. 

\begin{figure*}[t]
     \centering
         \includegraphics[width=3.5in,height=2.5in]{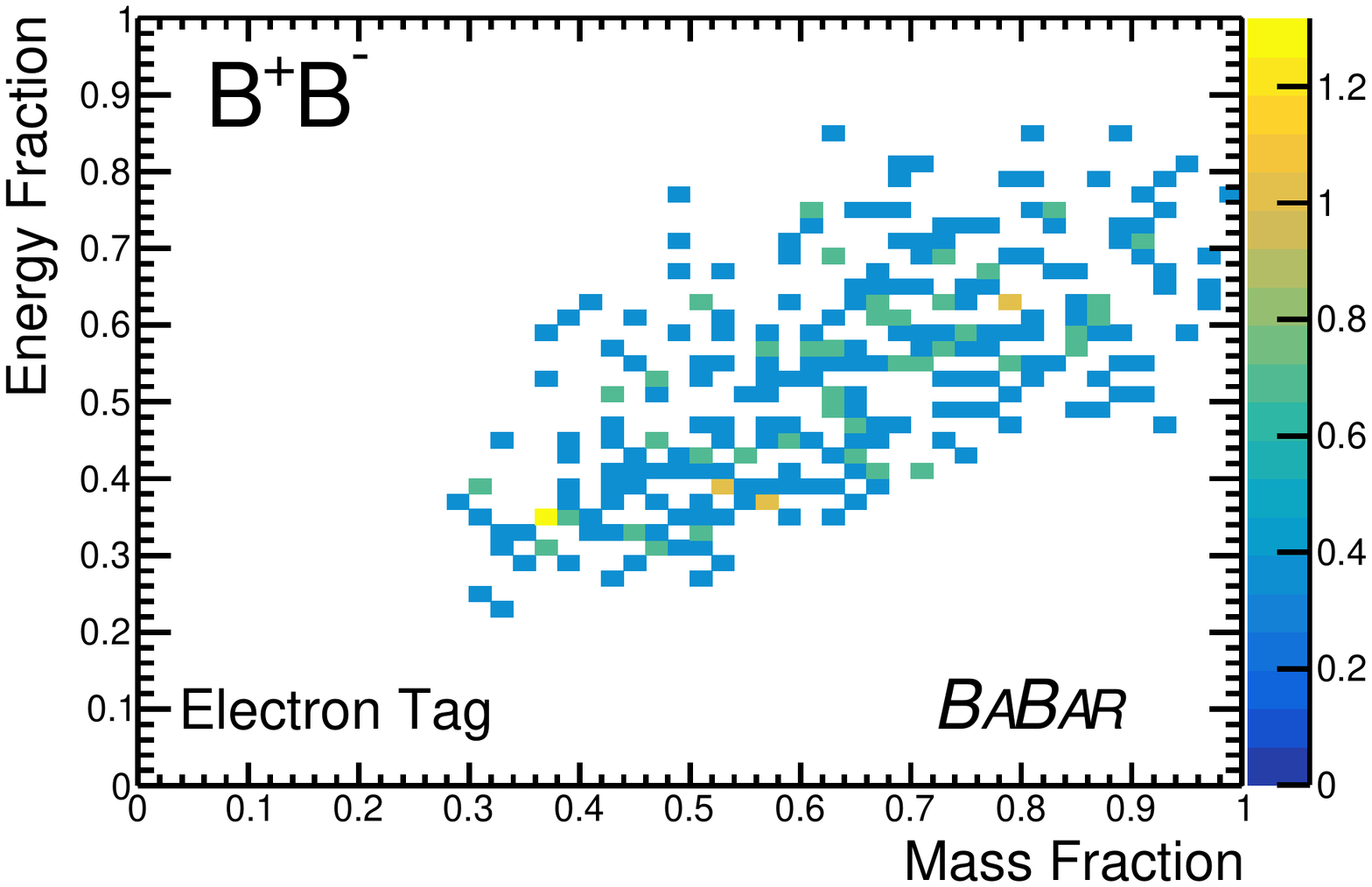}
         \includegraphics[width=3.5in,height=2.5in]{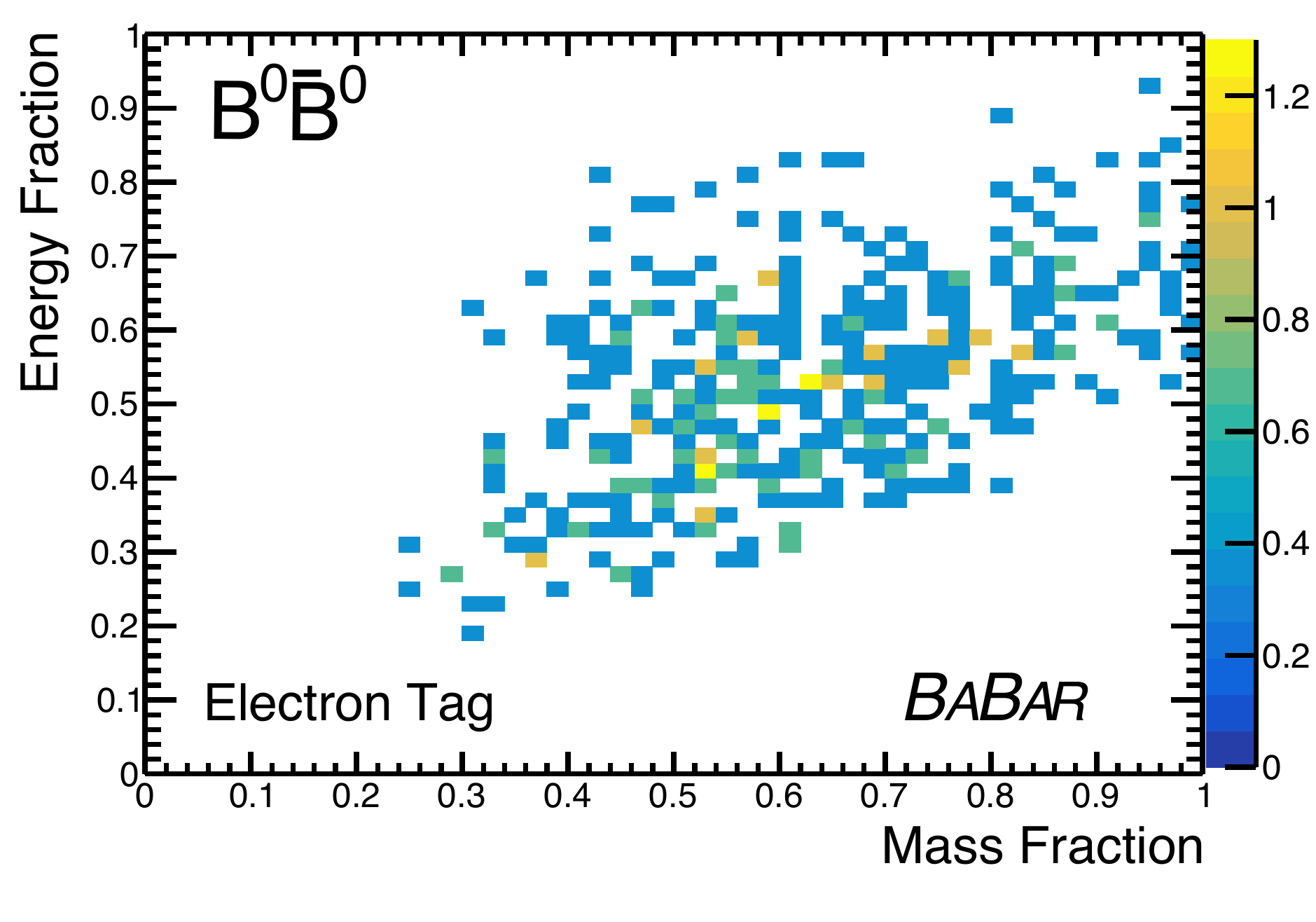}
         \includegraphics[width=3.5in,height=2.5in]{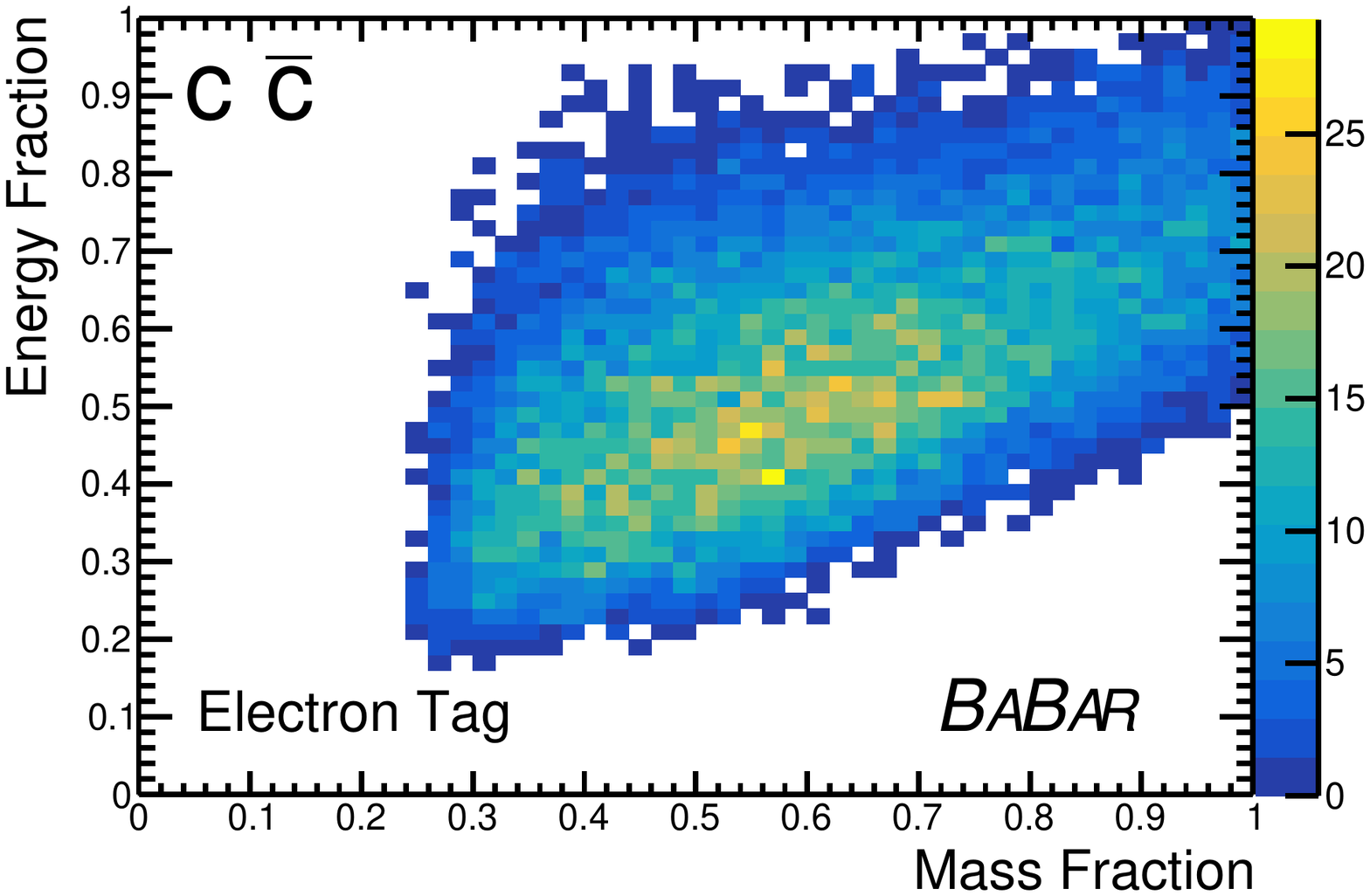}
         \includegraphics[width=3.5in,height=2.5in]{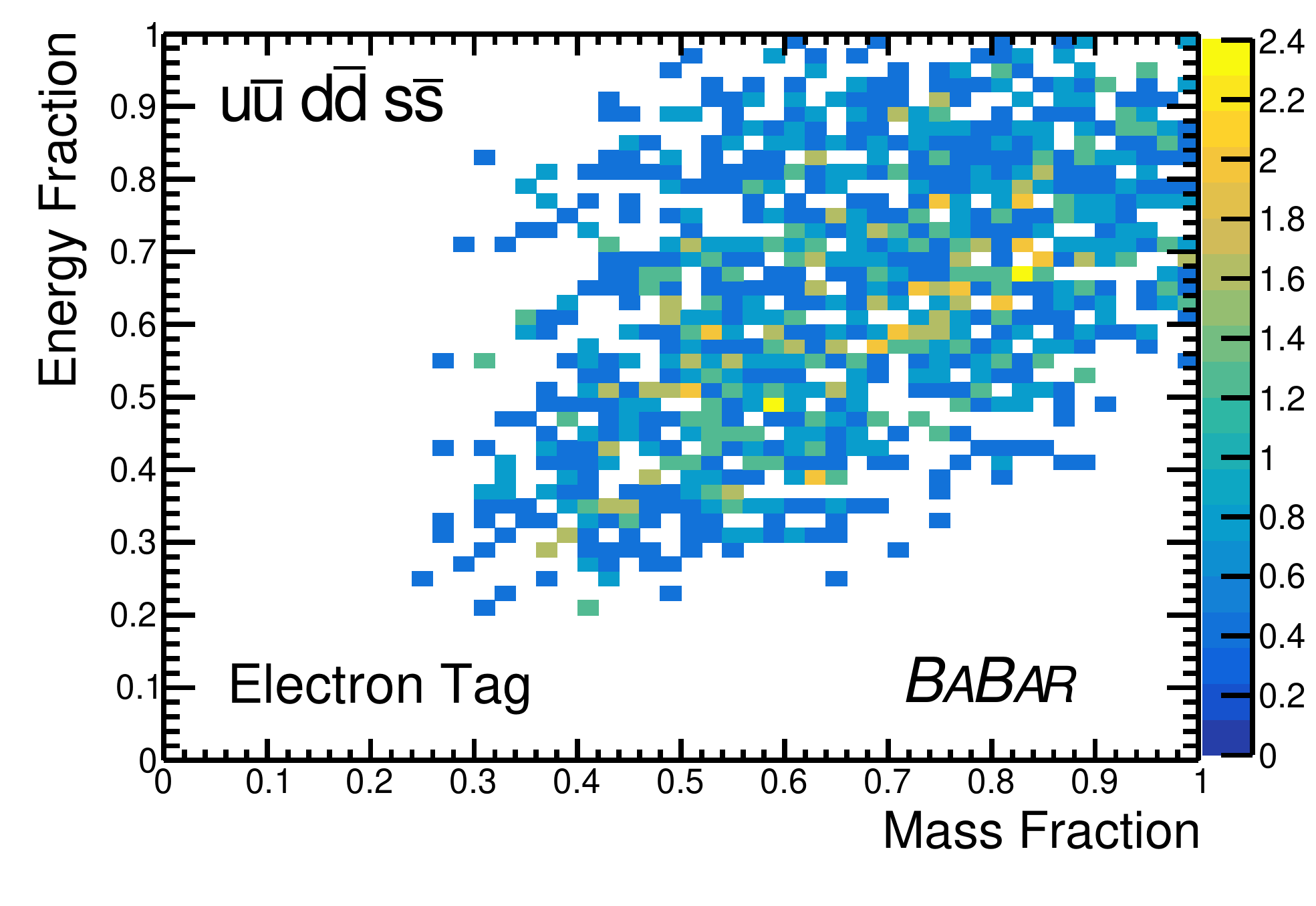}
        \caption{\textbf{Electron Tag (non-$\tau$)}: 2D templates showing reconstructed invariant-mass and energy (as fraction of incoming $\tau$ mass and
energy) for SM non-$\tau$ background processes, the $\mu^{+}\mu^{-}$ background is not shown since the yield is zero.}
        \label{fig:norm_bkg_elec}
\end{figure*}

\begin{figure*}[t]
    
         \includegraphics[width=3.5in,height=2.5in]{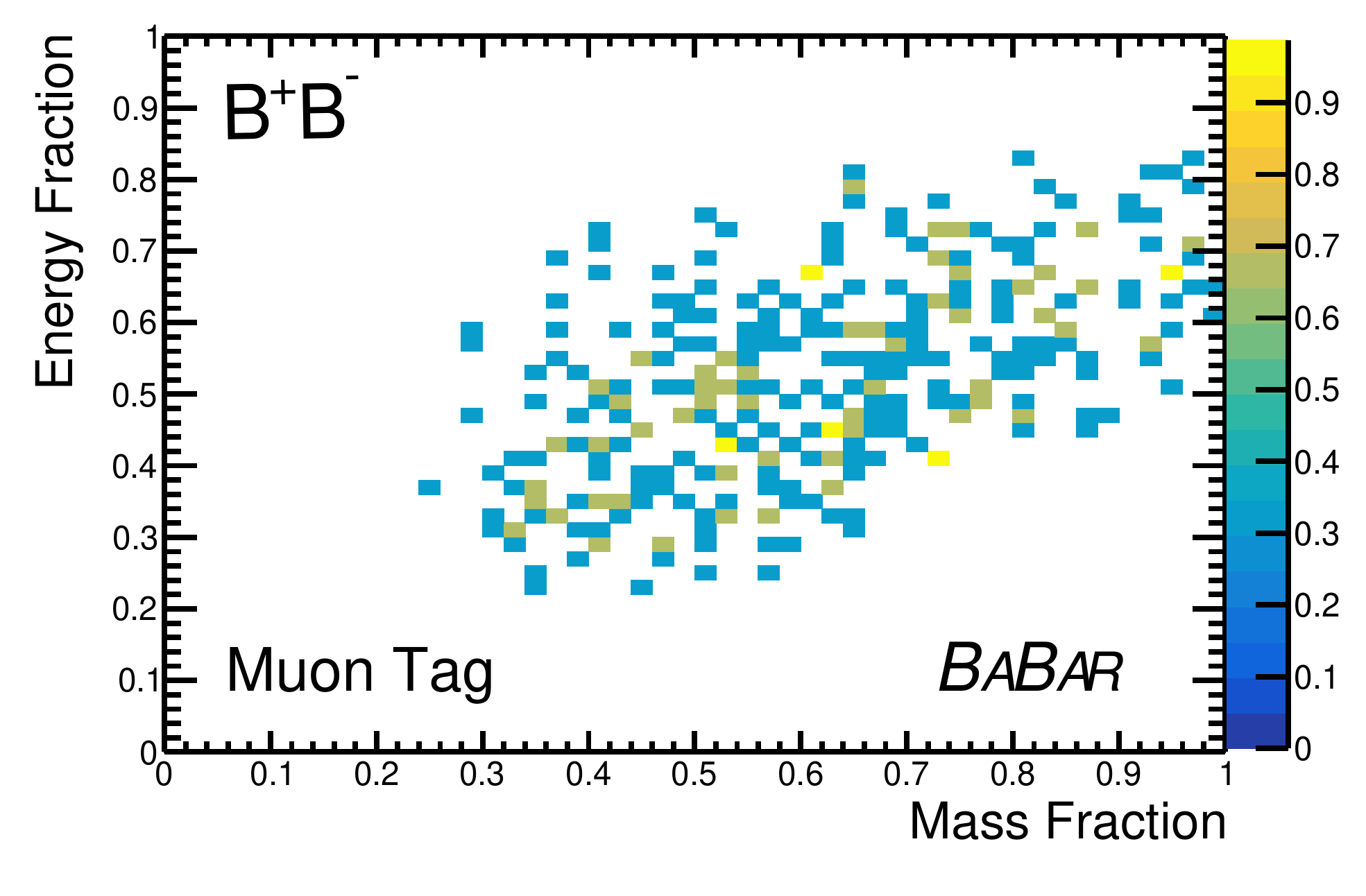}
         \includegraphics[width=3.5in,height=2.5in]{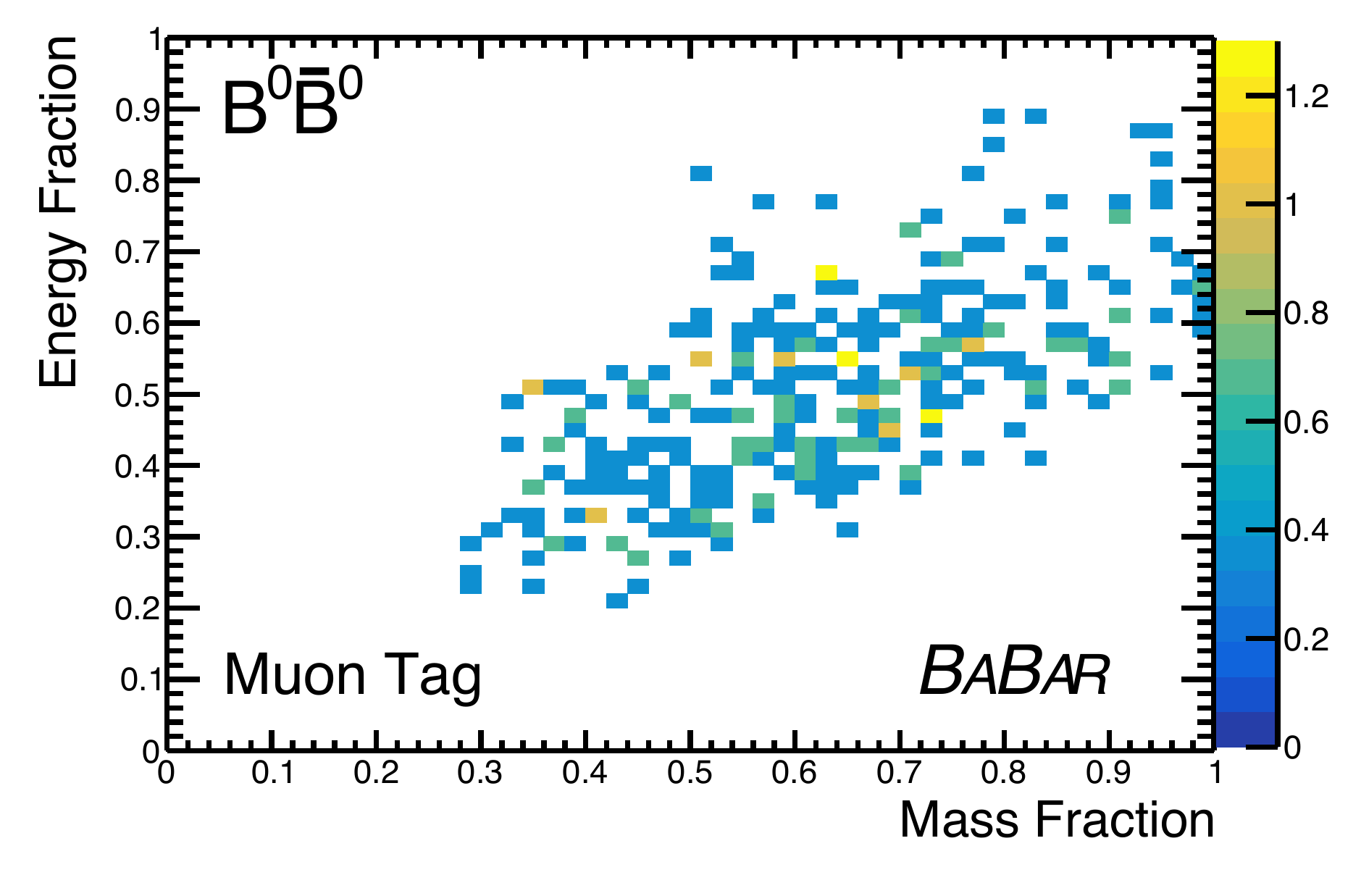}
         \includegraphics[width=3.5in,height=2.5in]{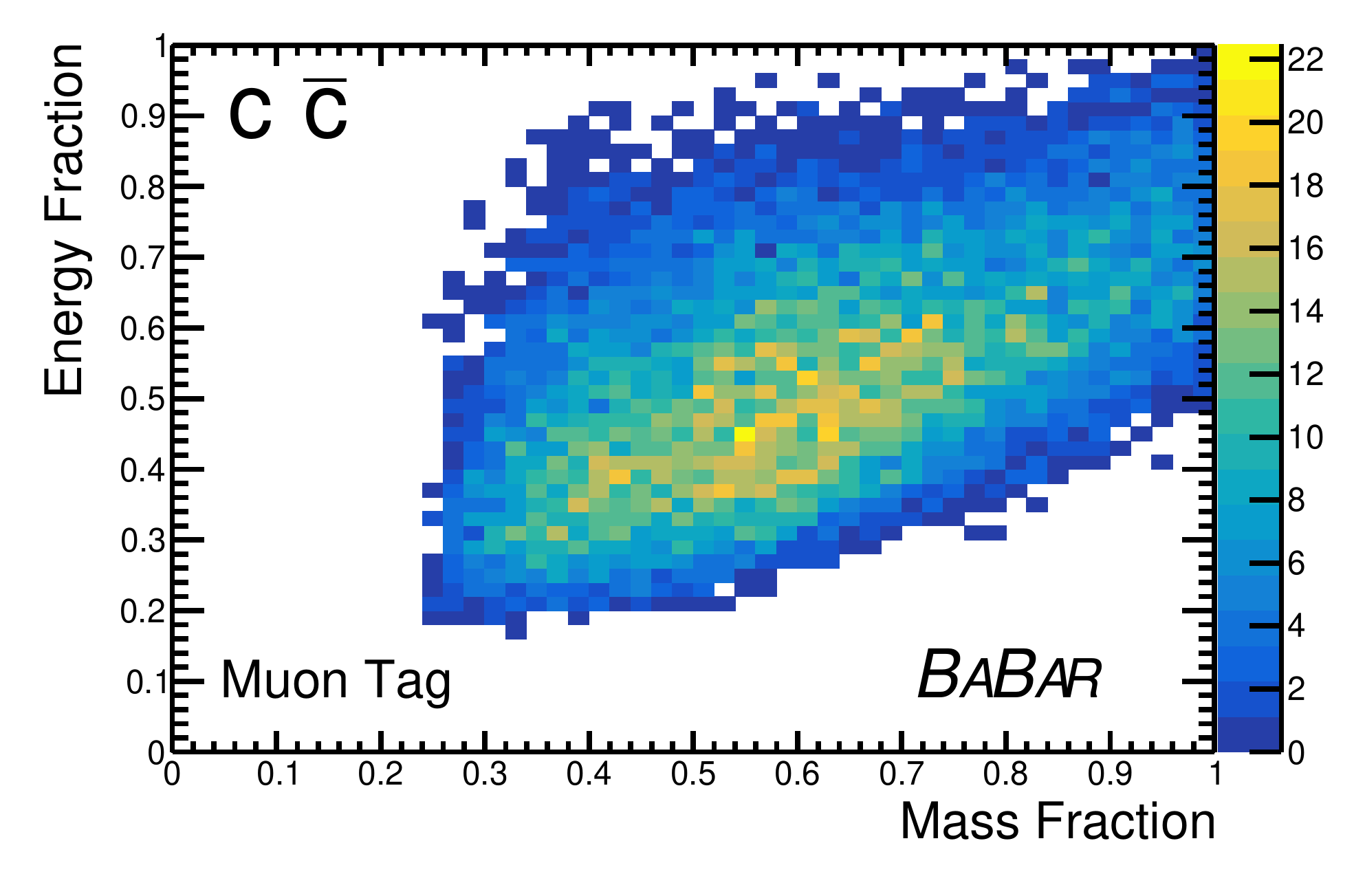}
         \includegraphics[width=3.5in,height=2.5in]{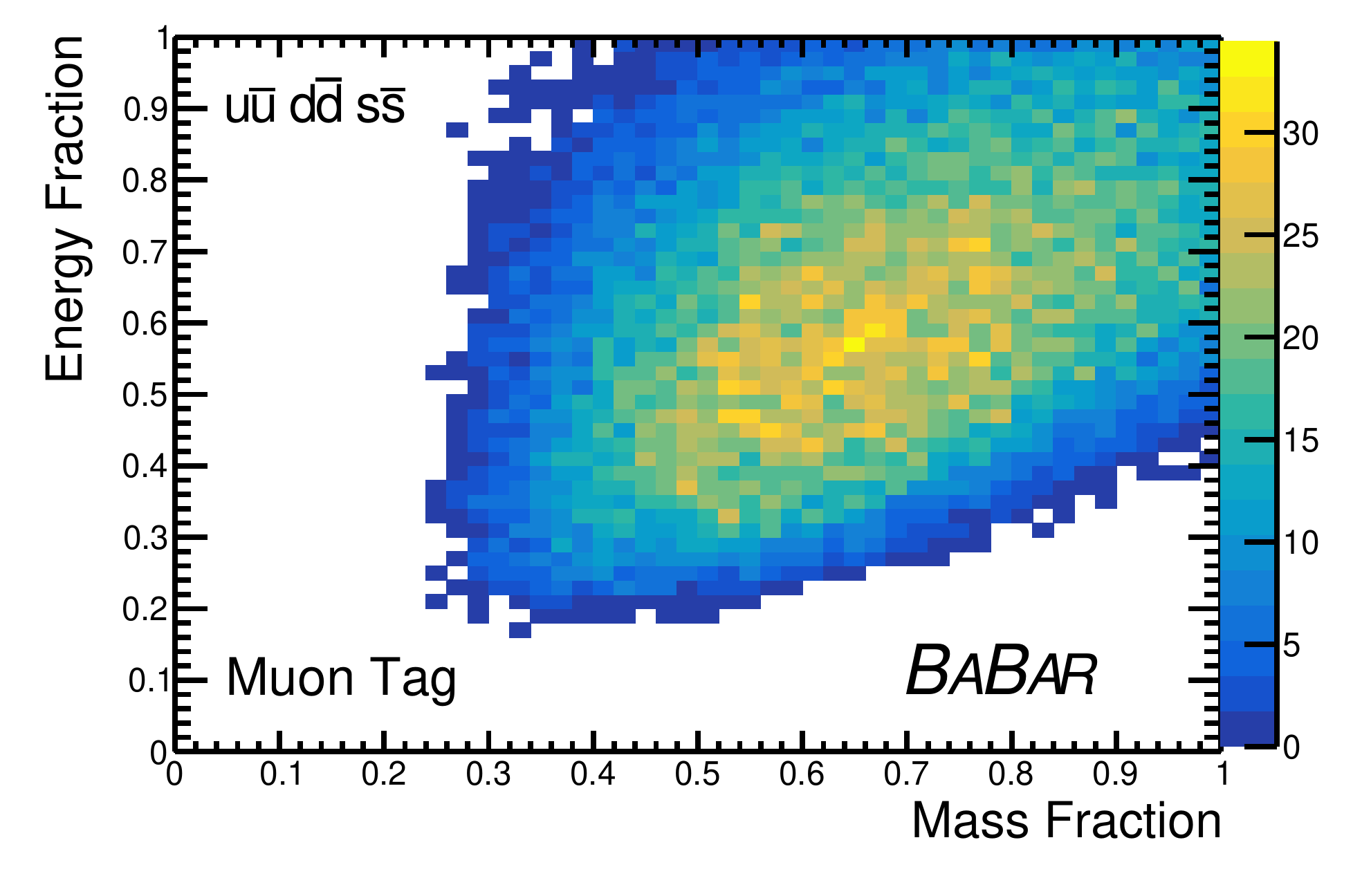}
         \includegraphics[width=3.5in,height=2.5in]{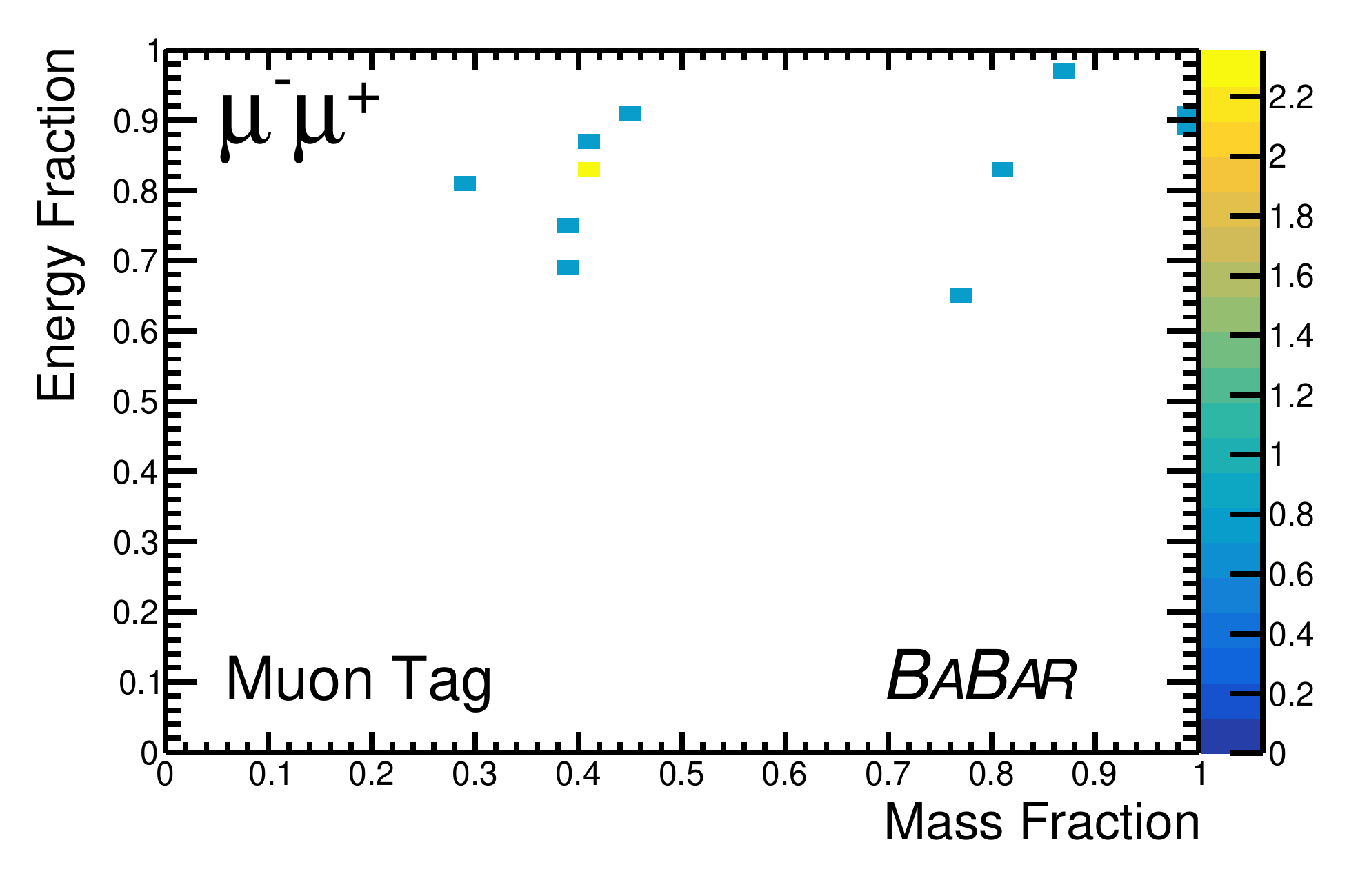}
        \caption{\textbf{Muon Tag (non-$\tau$)}: 2D templates showing reconstructed invariant-mass and energy (as fraction of incoming $\tau$ mass and
energy) for SM non-$\tau$ background processes.}
        \label{fig:norm_bkg_mu}
\end{figure*}

\begin{figure}[t]
         \includegraphics[width=3in,height=2in]{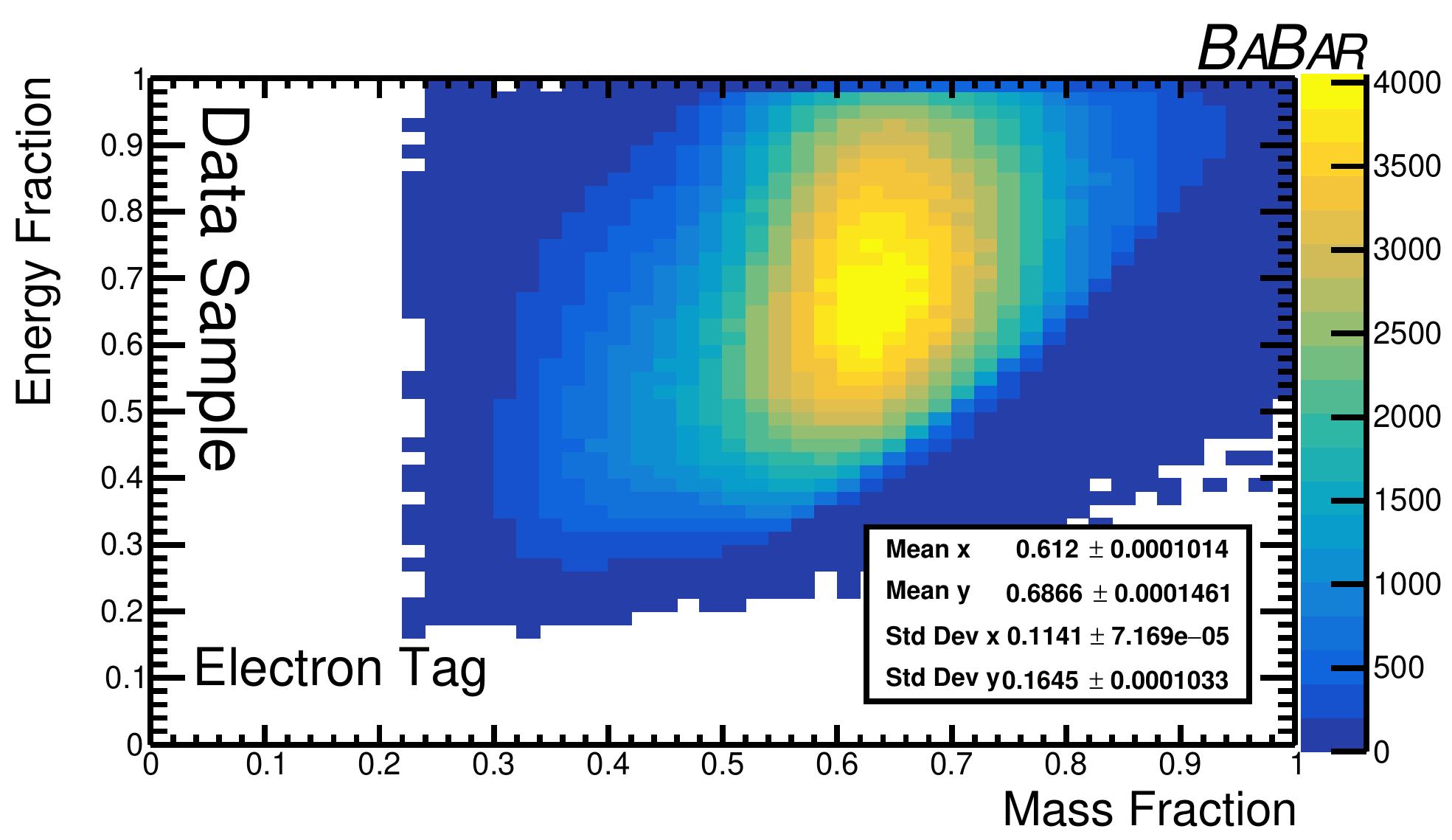}
         \includegraphics[width=3in,height=2in]{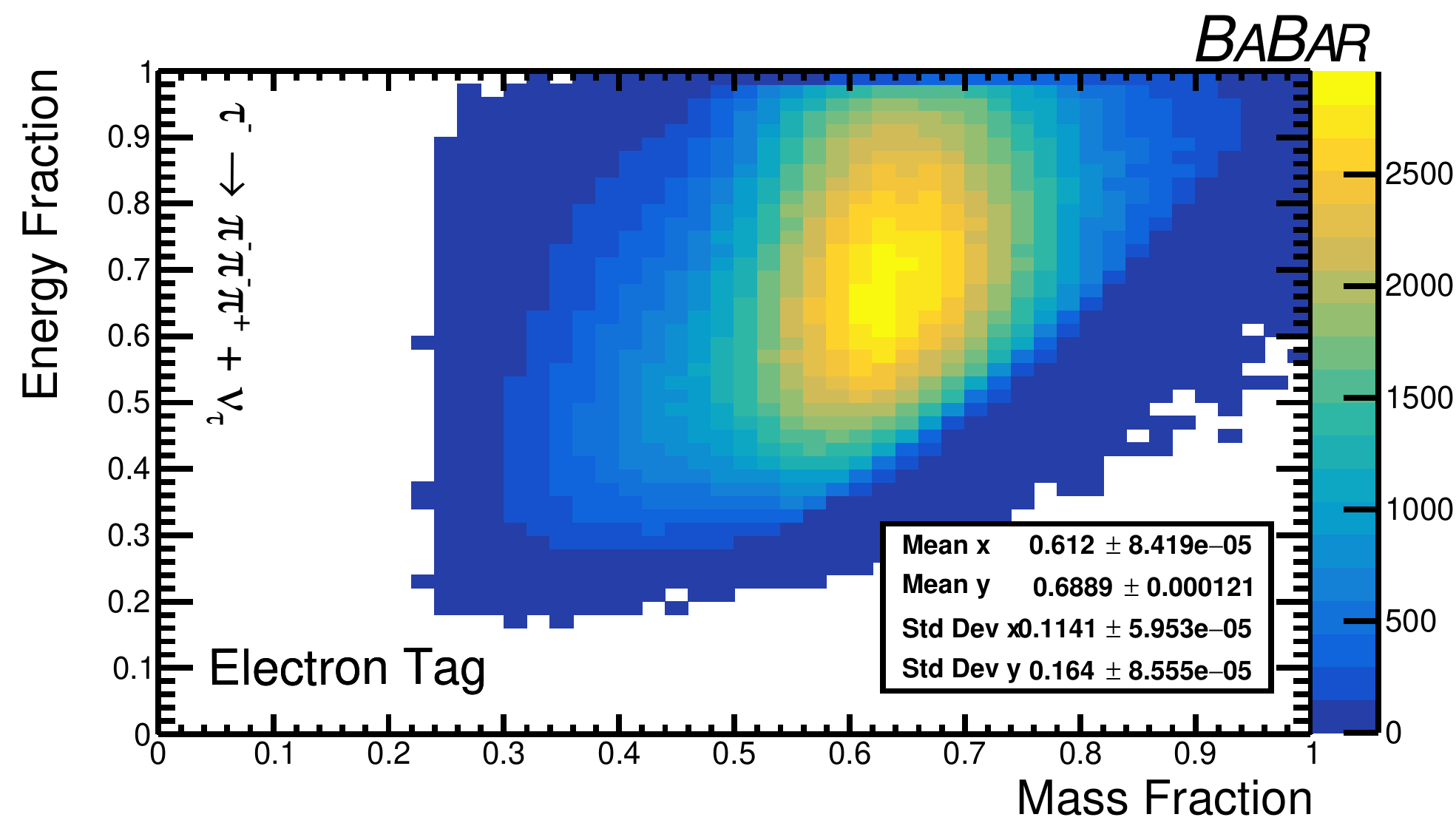}
         \includegraphics[width=3in,height=2in]{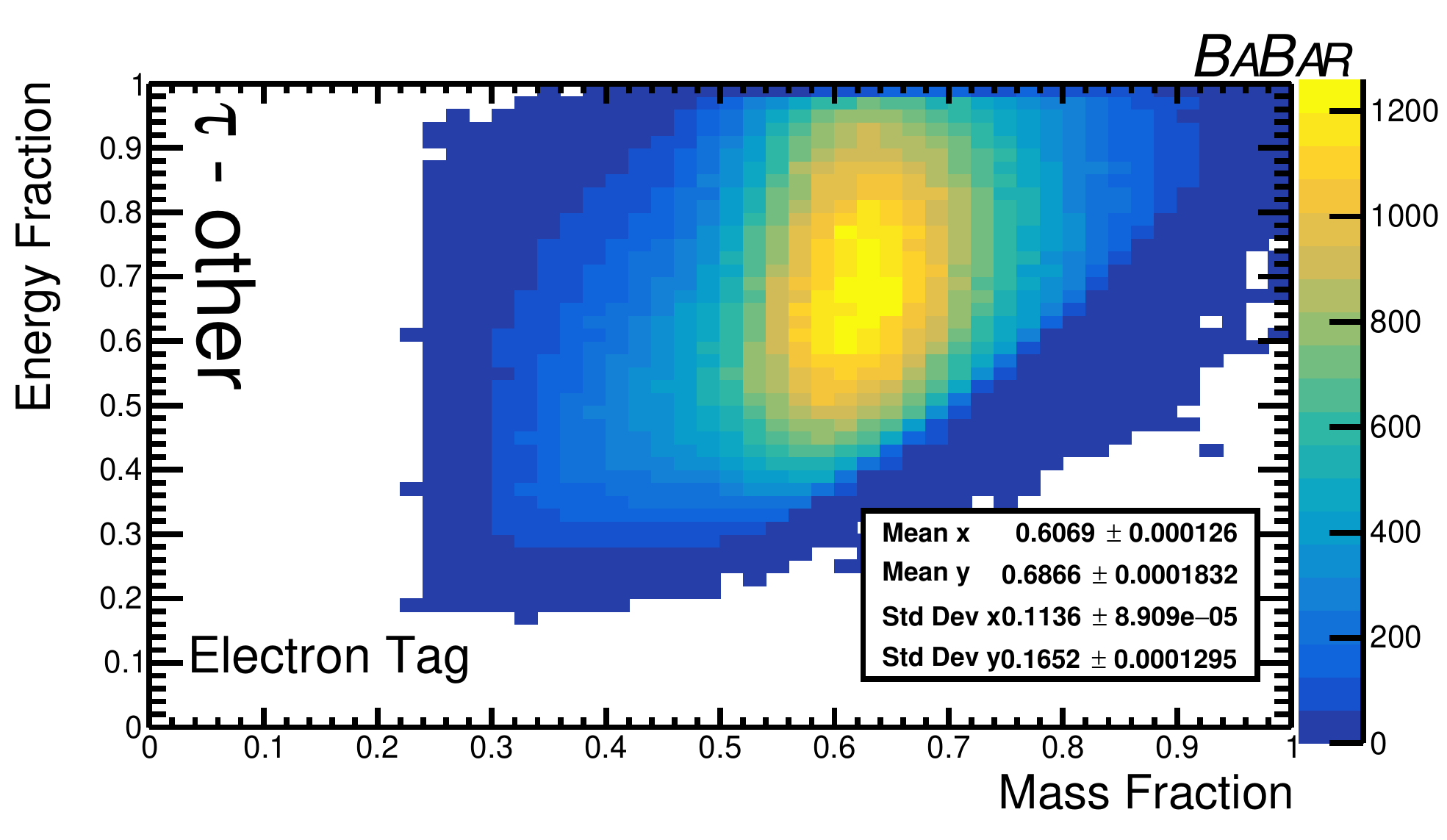}
        \caption{\textbf{Electron Tag}: 2D Templates of reconstructed hadronic invariant-mass and energy ($m_{h},E_{h}$) (as fraction of incoming $\tau$ mass and
energy) for data (top), SM $\tau$ MC template for $\tau^{-} \rightarrow \pi^{-}\pi^{-}\pi^{+}\nu_{\tau}$ (middle) and all other $\tau$ decays (bottom). }
        \label{fig:norm_tau_elec}
\end{figure}

\begin{figure}[t]
         \includegraphics[width=3in,height=2in]{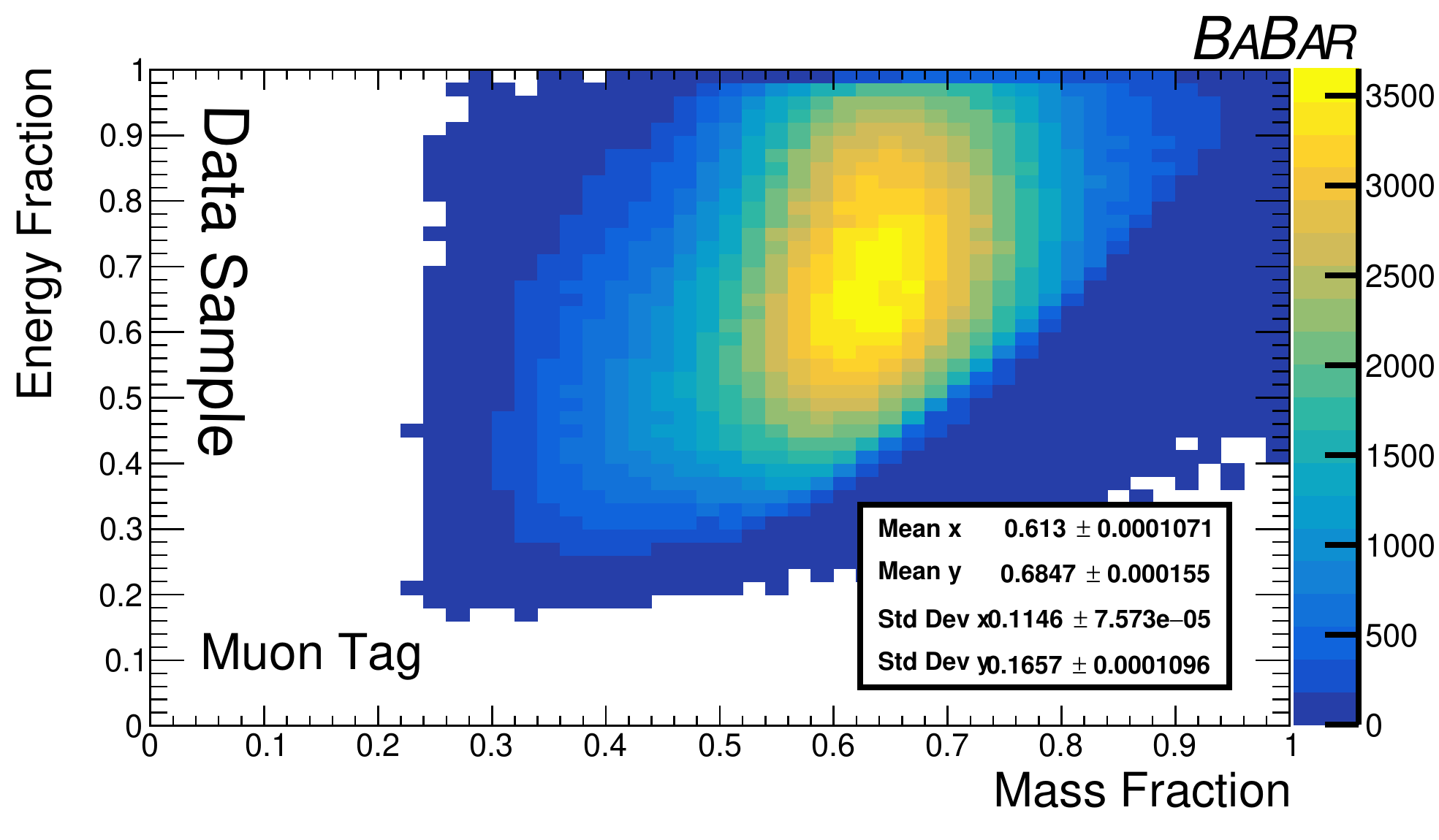}
         \includegraphics[width=3in,height=2in]{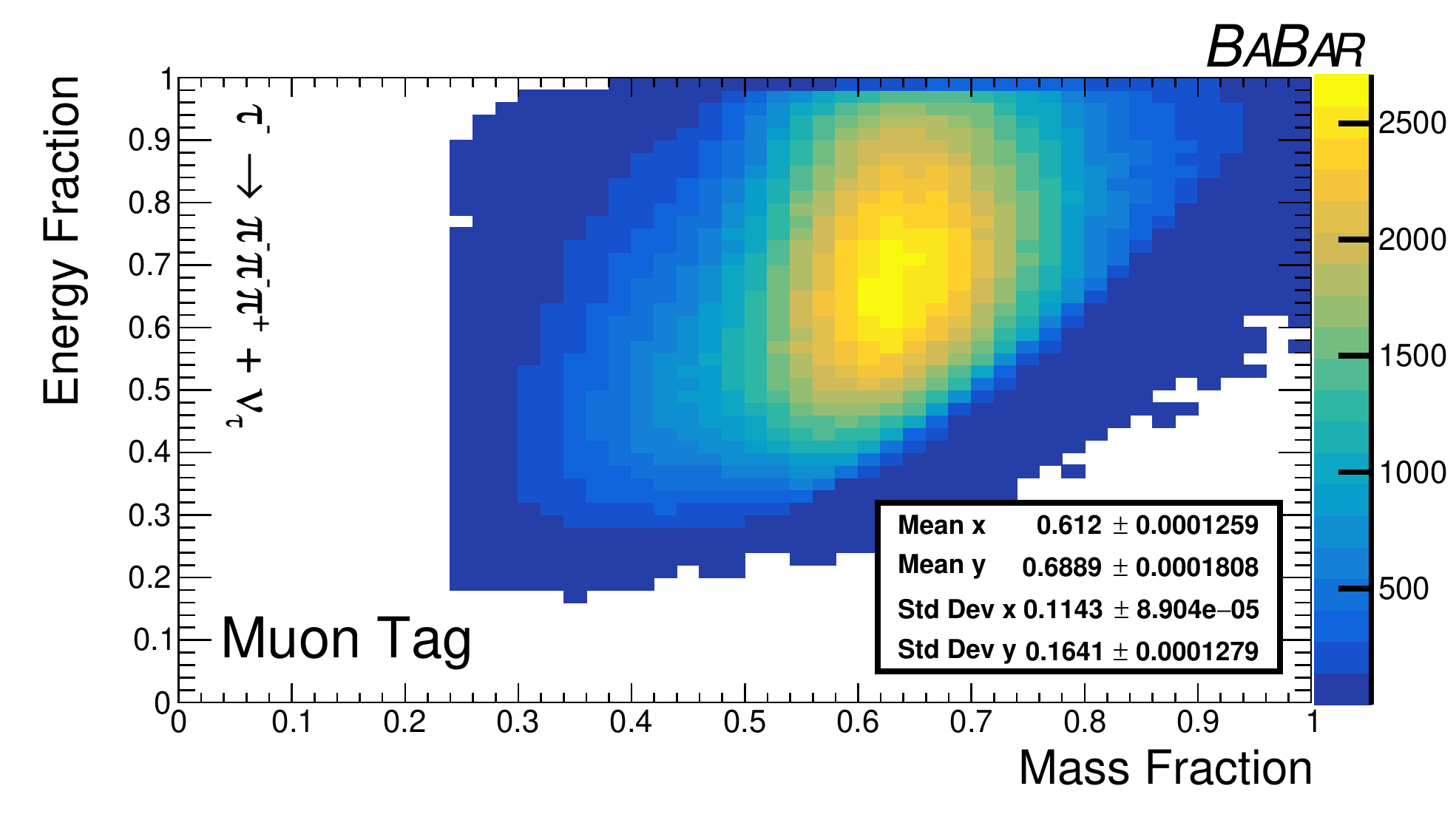}
         \includegraphics[width=3in,height=2in]{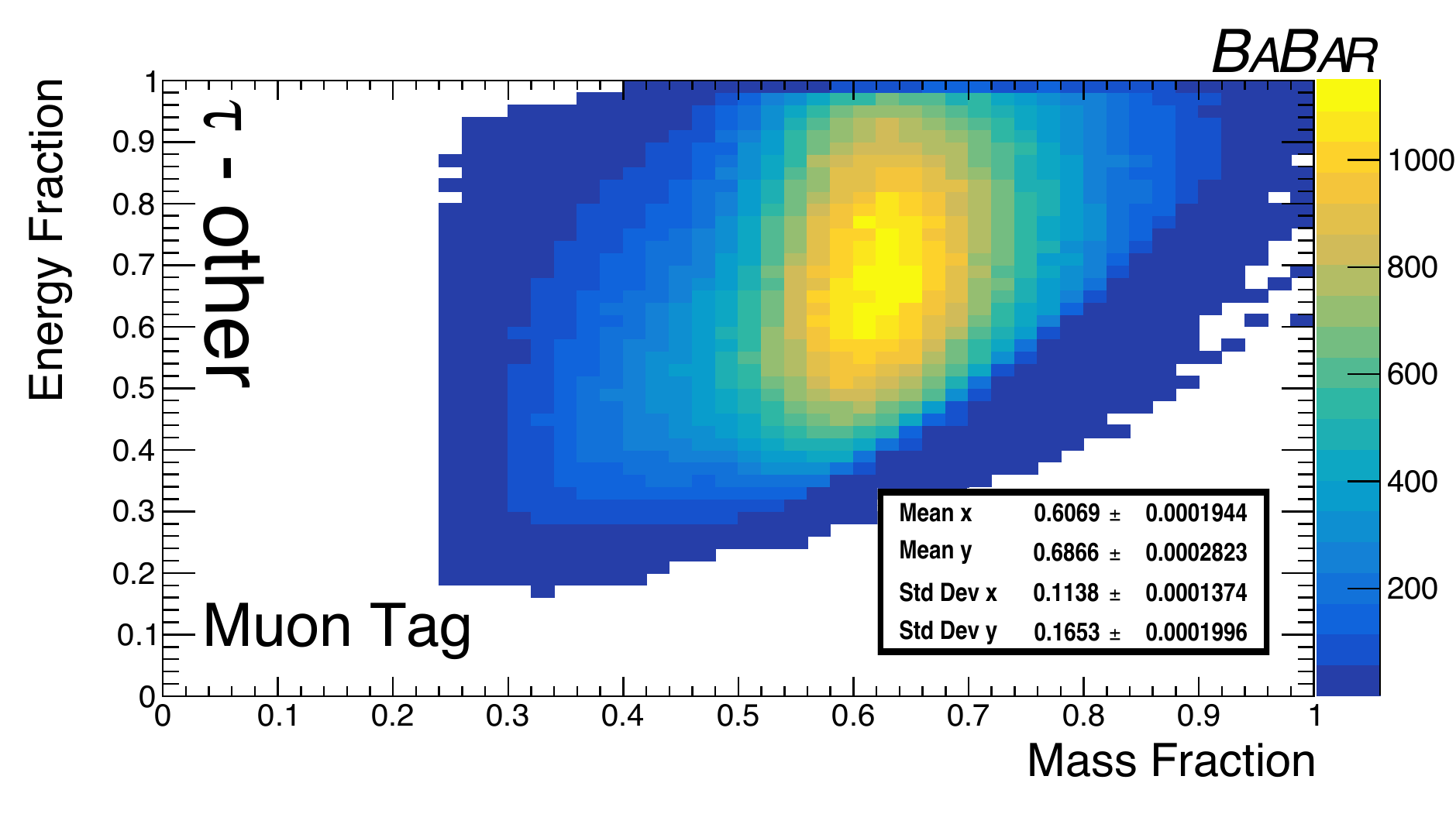}
        \caption{\textbf{Muon Tag}: 2D Templates of reconstructed hadronic invariant-mass and energy ($m_{h},E_{h}$) (as fraction of incoming $\tau$ mass and
energy) for data (top), SM $\tau$ MC template for $\tau^{-} \rightarrow \pi^{-}\pi^{-}\pi^{+}\nu_{\tau}$ (middle) and all other $\tau$ decays (bottom). }
        \label{fig:norm_tau_mu}
\end{figure}

Figures ~\ref{fig:norm_bkg_elec} and ~\ref{fig:norm_tau_elec} and  ~\ref{fig:norm_bkg_mu} and ~\ref{fig:norm_tau_mu} show the 2D template histograms for all the processes in the electron and muon tag, respectively (negative channel only, for simplicity). The largest background source comes from the channels which include 3-prong pionic decays, accompanied by a neutral pion, as outlined in Tables~\ref{table:bkg_pos} -~\ref{table:bkg_neg}. The main difference being that the mean is lowered, there is missing mass, corresponding to neutral pions. This must be well-accounted for to not misidentify this as a HNL of mass $\sim$ $m_{\pi^{0}}$. Uncertainties in modeling of these non-signal $\tau$ channels are also included in the background uncertainty shown, and will be discussed in Sec.~\ref{sec:tau_shape}.

\section{\label{sec:likelihood}Binned Likelihood Function}

A binned likelihood approach is taken to place limits on the parameter of interest $|U_{\tau 4}|^{2}$, the mixing parameter between the SM $\tau$ neutrino and the HNL. It is assumed that the contents of a given bin, $i,j$, in the $(m_{h},E_{h})$ data histogram is distributed as a Poisson distribution and may contain events emanating from any of the SM process, and potentially HNL signal events. The number of expected events reconstructed in a given bin ($ \nu^{\text{reco}}_{\text{obs},ij}$) may be written as:

\begin{equation} \nu^{\text{reco}}_{\text{obs},ij}= (\nu_{\text{\tiny HNL},ij} + \nu_{\tau-SM,ij} + \nu_{\tau-\text{other},ij} + \nu_{\text{non}-\tau,ij})^{\text{reco}},\end{equation}
where $\nu_{\text{\tiny HNL},ij}$ is the expected number of signal events, $\nu_{\tau-\text{SM},ij}$ represents the expected number of events from the SM $\tau^{-} \rightarrow \pi^{-}\pi^{-}\pi^{+}\nu_{\tau}$ decay, and $\nu_{\tau-\text{other},ij}$ and $\nu_{\text{non}-\tau,ij}$ are the expected SM other $\tau$ and non-$\tau$ backgrounds, respectively. The values of $\nu_{\tau-\text{SM},ij}$, $\nu_{\tau-\text{other},ij}$ and $\nu_{\text{non}-\tau,ij}$ are inferred from MC simulation;  $\nu_{\text{\tiny HNL}}$ must be estimated using our 2D template histograms for a given mass.

Denoting the number of generated $\tau$-lepton events in the sample from a specific tag as:

\begin{equation}
    N_{\tau, \text{gen}} =  \mathcal{L}_{\text{int}} \cdot \sigma(ee\rightarrow \tau\tau) \cdot   BR(\text{3-prong}) \cdot BR(\text{1-prong}),
\end{equation}
the numbers of expected signal, SM $\tau$ 3-prong, and other background events are written more simply as:

\begin{equation}
    \hat{\nu}_{\text{\tiny HNL},ij} = n^{\text{reco}}_{\text{\tiny HNL},ij} =   N_{\tau, \text{gen}}\cdot (|U_{\tau 4}|^{2}) \cdot p_{\text{\tiny HNL}, ij},
\end{equation}
\\
\begin{equation}
    \hat{\nu}_{\tau-\text{SM},ij} = n^{\text{reco}}_{\tau-SM,ij} =  N_{\tau, \text{gen}}\cdot (1-|U_{\tau 4}|^{2}) \cdot p_{\tau-SM, ij} ,
\end{equation}

and
\begin{equation}
    \hat{\nu}_{\text{BKG},ij} = n^{\text{reco}}_{BKG,ij} = n^{\text{reco}}_{\tau-\text{other},ij} + n^{\text{reco}}_{\text{non}-\tau,ij},
\end{equation}
where the final term ($ \hat{\nu}_{\text{BKG},ij}$) is a combination of all the backgrounds not associated with a $\tau^{-} \rightarrow \pi^{-}\pi^{-}\pi^{+}\nu_{\tau}$ channel, calculated from MC. The $p_{ij}$ terms are the products of the reconstruction efficiency ($\epsilon^{\text{reco}}_{ij}$ - the probability that, given an event took place, that the reconstruction algorithms find it), selection efficiency ($\epsilon^{\text{sel}}_{ij}$ - the probability that the reconstructed event passes the event selection in this analysis), and $p^{\text{shape}}_{ij}$ (the fraction of total histogram reconstructed and selected events present in the $ij$-th bin) for each process.

The likelihood to observe the selected candidates in all the $(m_{h}, E_{h})$  bins is the product of the Poisson probability to observe the selected events in each bin:

\begin{widetext}

\begin{equation} \mathcal{L} = \prod_{ij}  f(n_{ij};n_{\text{obs}},\vec{\theta}) = \prod_{ij} \frac{(\nu_{\text{\tiny HNL}}+\nu_{\tau-\text{SM}}+\nu_{\text{BKG}})_{ij}^{(n_{\text{obs}})_{ij}}e^{-(\nu_{\text{\tiny HNL}} + \nu_{\text{BKG}} + \nu_{\tau-\text{SM}})_{ij} }}{(n_{\text{obs}})_{ij}!}  \times \prod_{k} f(\theta_{k},\tilde{\theta}_{k}) ,  \end{equation}

\end{widetext}
where $n_{\text{obs}}$ is the number of events observed in the data and $\vec{\theta}$ describes a number of nuisance parameters corresponding to the yield uncertainties outlined in Sec.~\ref{sec:syserrors}. The final product in this expression represents the product of the nuisance parameters. Each parameter, $k$, will be modeled using a Gaussian probability density function: $f(\theta_{k},\tilde{\theta}_{k})$. Each $f(\theta_{k},\tilde{\theta}_{k}) $ term represents the probability for the true value of a nuisance parameter to be equal to $\theta_{k}$, given that the best estimate of the parameter is $\tilde{\theta}_{k}$, which is determined using the methods outlined in Sec.~\ref{sec:syserrors}.

Substituting in the expressions for the estimators of the expected yields gives:

\begin{widetext}

\begin{align*}
\mathcal{L} =   \prod_{\text{charge}}^{+ -} \bigg ( \prod_{\text{channel}}^{e \mu } \bigg (  \prod^{ij}_{\text{bin}}  \bigg ( \frac{1}{n_{\text{obs},ij}!}\bigg[ N_{\tau,\text{gen}}\cdot |U_{\tau 4}|^{2} \cdot p_{\text{\tiny HNL}, ij}+N_{\tau,\text{gen}}\cdot (1-|U_{\tau 4}|^{2}) \cdot p_{\tau-\text{SM}, ij} + n^{\text{reco}}_{BKG,ij}\bigg]^{(n_{\text{obs}})_{ij}} \times
\end{align*}
\begin{equation}
exp\bigg[-(N_{\tau,\text{gen}}\cdot |U_{\tau 4}|^{2} \cdot p_{HNL, ij} + N_{\tau,\text{gen}}\cdot (1-|U_{\tau 4}|^{2}) \cdot p_{\tau-SM, ij}+ n^{\text{reco}}_{BKG,ij}) \bigg] \bigg )_{\text{bin}}
\times  \prod_{k} f(\theta_{k},\tilde{\theta}_{k}) \bigg )_{\text{channel}} \bigg )_{\text{charge}}.
\label{eq:lik}
\end{equation}
\end{widetext}
The expression involves a product over all bins, $ij$, over the two 1-prong channels, and over both possible $\tau$-lepton charges ($\pm$). 

In this analysis it is assumed that the \babar~ modeling and reconstruction software offers a realistic representation of the physics processes, the experimental environment and the response of the detectors to the data. Any known cause of a discrepancy between the data and MC must be characterized as an uncertainty. 

Reference \cite{conway} gives an overview of the process of incorporating nuisance parameters into a binned Poisson likelihood such as that presented. The expected number of signal and background events are functions of a set of nuisance parameters ($\vec{\theta}$). These represent the systematic yield uncertainties to be outlined in Sec.~\ref{sec:syserrors}. Incorporation of shape uncertainties is described in  Secs~\ref{a1}-~\ref{sec:tau_shape}. 

Two hypotheses can be proposed regarding the content of the data:

\begin{itemize}
\item $ H_{0}$:  where the signal is present at some level, $|U_{\tau 4}|^{2}_{0}$,
\item  $ H_{1} $: where the data is a mixture of background + signal, with small signal present.
\end{itemize}

A likelihood ratio test statistic ($LR$) is used to test hypothesis $H_{0}$ against $H_{1}$:

\begin{equation} LR  = \frac{\mathcal{L}_{H_{0}}(|U_{\tau 4}|_{0}^{2};\hat{\hat{\theta}}_{0},\text{data})}{\mathcal{L}_{H_{1}}(|\hat{U}_{\tau 4}|^{2};\hat{\theta},\text{data}) }, \end{equation}
where $\mathcal{L}$ in both numerator and denominator describes the maximized likelihood, for two instances. The denominator is the maximized (unconditional) likelihood giving the maximum likelihood estimator of $|U_{\tau 4}|^{2}$ and the set of nuisance parameters ($\hat{\theta}$); $\hat{\theta}$ is a vector of nuisance parameters which maximize the likelihood. In the numerator the nuisance parameters are maximized for a given value of $|U_{\tau 4}|^{2}$ i.e. it is the conditional maximum-likelihood. The ratio, $LR$, is consequently a function of $|U_{\tau 4}|^{2}$ through the numerator. It must be noted that the numerator denotes the hypothesis for any given value of $|U_{\tau 4}|^{2}$ (including the background only case i.e. $|U_{\tau 4}|^{2}=0$). Reference \cite{Cowan:2010js} provides more details on likelihood-based tests.

A test statistic, $q$, can be defined as:

\begin{equation}
    q = -2 \text{ln} \bigg (  \frac{\mathcal{L}_{H_{0}}(|U_{\tau 4}|_{0}^{2};\hat{\hat{\theta}}_{0},\text{data})}{\mathcal{L}_{H_{1}}(|\hat{U}_{\tau 4}|^{2};\hat{\theta},\text{data}) } \bigg ) = -2\text{ln}(\Delta \mathcal{L}).
\end{equation}

Using Wilk's theorem \cite{wilks}, $q$ asymptotically approaches a $\chi^{2}$ distribution under $H_{0}$. To find a $100(1 - \alpha)\%$ confidence interval we move to the left and to the right of the minimum value of $q$ to find the points where the function increases by the $\alpha$ percentile of a $\chi^{2}$ distribution with a number degrees of freedom equal to the number of parameters. 

\section{\label{sec:syserrors} Systematic Uncertainties}

There are two types of systematic uncertainty which must be accounted for:

\begin{enumerate}
    \item \textbf{Normalization Uncertainties} which affect the overall expected yield of a particular background or signal events. These uncertainties affect all bins uniformly.
    \item \textbf{Shape Uncertainties} which affect the shape of either the background or signal in the template histograms (in the ($m_{h},E_{h} $) plane). These affect the signal/background yield in a specific bins differently and may mean that template distribution shapes are shifted non-uniformly.
\end{enumerate}

Since the expected background contributions are estimated using MC simulations, any uncertainty on that MC can affect how accurately it represents the real data. Therefore, in the following discussion emphasis is placed on quantifying sources of  deviation between the MC and data and the impact of these uncertainties on the final result.

\subsection{Normalization Uncertainties}

Normalization uncertainties are incorporated into this analysis as nuisance parameters, parameterized as Gaussian functions, unless otherwise stated: 

\subsubsection{Luminosity}

Uncertainty on the luminosity are calculated from experimental systematic uncertainties on the Bhabha and muon-pair  selections used in luminosity determination. The uncertainty is run-dependent and averages at $0.44 \%$\cite{BaBar}.

\subsubsection{Cross-section}
There is an uncertainty associated with the $\tau$-pair production cross-section $\sigma(e^{+} e^{-} \rightarrow \tau^{+} \tau^{-}) = 0.919 \pm 0.003$ nb. A relative uncertainty of 0.3$\%$ is assigned.

\subsubsection{Uncertainties in $\tau$ Decay Mode Branching Fractions}

The decay branching fractions for the $\tau$-lepton decay modes used within TAUOLA were updated to reflect the current best-fit values listed in Ref. \cite{newpdg}. Each of these is an average from several experimental results and each has an associated error. 

This uncertainty will affect the yield of each $\tau$ related channel across all bins i.e. it is a scaling factor. It should not affect the shape of the template histograms. To account for these uncertainties the yields of each of these backgrounds are varied by $\pm 1 \sigma$ on the listed percentage and the final analysis is repeated. The overall variation is included in the systematic uncertainty on the final result. The systematic uncertainty on the final result, due to these branching fraction uncertainties is very small $\ll 0.1 \%$.

\subsubsection{PID Efficiencies}
\label{pid_sys}

In this analysis we rely on the \babar~ PID algorithms \cite{BaBar} to accurately select leptonic 1-prong and hadronic 3-prong events in both the MC and data samples. 

The electron selection algorithm has an efficiency of above 90$\%$ and a pion mis-identification rate between $0.05-0.01\%$. Efficiencies for negative and positive tracks were equivalent.  We estimate a systematic uncertainty of 2$\%$ to be consistent with the
observed variations of data and MC.

This muon selection algorithm has an efficiency of $\sim 80 \%$ and a mis-identification rate of $>1.5\%$. Only small differences between MC and data was observed in the test sample, therefore, a systematic uncertainty of 1$\%$ is assigned.

For the 3-prong hadronic channel the pion selection algorithm has an efficiency of above $\sim 85\%$, a kaon mis-identification rate of $\sim 2 - 4 \%$, an electron mis-identification rate of   $\sim 5\%$, and a muon mis-identification rate of $\sim 20\%$. A systematic uncertainty of 3$\%$ is assigned.

All PID uncertainties are included as nuisance parameters.

\subsubsection{Bin Size}

Altering the number of bins in the 2D template histograms by double or half the number of bins had a small effect of 0.2 $\%$ on the end result.

\subsubsection{Other Systematic Uncertainties}

The following systematic uncertainties were ignored in the final study:

\begin{enumerate}
    \item \textbf{Tracking Efficiency}

The tracking efficiency has been measured in a general purpose ~\babar~ collaboration study using control samples of charged tracks for both MC and data for all run periods. The total uncertainty (average for all runs) is 0.5$\%$. Since the simulation and data samples efficiency are consistent, within uncertainties, no efficiency correction is applied.

 \item \textbf{Trigger and Filter Efficiency}

Corrections to account for differences in the filter and trigger efficiency are found to be negligible.

 \item \textbf{Modeling of Detector Response}
\label{response}

Deviations between data and MC due to detector response modeling are negligible. 

 \item \textbf{Beam Energy}

There is a systematic uncertainty associated with the uncertainty on the beam collision energy. The systematic resulting from the CM energy scale and energy spread was determined as the maximum shift resulting from varying the CM energy in the MC by $\pm$ 2 MeV. This was found to have negligible effect on the final result.

 \item \textbf{Tau Mass}

The uncertainty on the $\tau$ mass ($m_{\tau}$  =  1776.99 $\pm$ 0.29 \mevcc) \cite{newpdg} has insignificant  effect on the results.

\item \textbf{$\tau$ Polarization}

No systematic effect is expected from $\tau$ polarization.

\end{enumerate}

\subsection{Shape Uncertainties}
Uncertainties on signal and background shapes require more careful analysis. These include:

\subsubsection{Uncertainties on non-\texorpdfstring{$\tau$}~ Backgrounds}

\begin{itemize}
\item  The side band region above the 3-prong invariant-mass requirement, $m_{123} > 1.77$ \gevcc~ is used to determine the deviation between data and the $q\bar{q}$  background distributions. The data were found to disagree with the MC in this region by $< 0.1 \%$.

\item The generic $\tau$-lepton MC does not contain Bhabha backgrounds. In this analysis all of the hadronic (3-prong) tracks were required to fail the electron PID selector. To estimate Bhabha contamination, this requirement was loosened to require only 0, 1 or 2 pions to fail the electron PID and then measuring the percent deviation between the data and MC in the region $ 0.6< 2p_{\text{CM}} \cdot c < 0.9 $,  the control region for the Bhabha background, where $p_{\text{CM}}$ is the momentum in the CM frame. Using the percentage changes as each requirement is removed one can extrapolate the expected percentage difference between MC and data for the case when the requirement is enforced for all 3 pions in the 3-prong side. The estimated contamination in the control region is $0.2\%$. 
\end{itemize}

Using these methods a total 0.3$\%$ uncertainty is assigned for these two non-$\tau$ backgrounds. This is included in the background uncertainty.

\subsubsection{Uncertainty on modeling of \texorpdfstring{$a_{1}$}~ resonance}
\label{a1}

For many hadronic $\tau$ decay channels the relative uncertainties from experimental results are large, and their modeling in TAUOLA is discussed in detail in Ref.~\cite{tauola}.

A $\tau$-lepton decay to an odd number of pions occurs almost exclusively through the axial-vector current, and the 3-prong pionic $\tau$-lepton decay is mediated by a $a_{1}(1260)$ meson with quantum numbers $J^{PC} = 1^{++}$ in 97$\%$ of cases. The $a_{1}(1260)$  resonance decays through the intermediate $\rho \pi$ state.

The experimental data on the $a_{1}(1260)$ may be grouped into two classes: hadronic production and  $\tau$-lepton decays. In the MC samples used in this analysis the PDG \cite{newpdg} average of $m_{a_{1}} = 1230 \pm 40$ \mevcc~ and Breit-Wigner averaged width of $\Gamma_{a_{1}} = 420 \pm 35 $ are used.  These averages include many different experiments, which look at both types of data. The uncertainty associated with the $a_{1}$ resonance represents the dominant contribution to the systematic error in our measurement.

In order to understand the effects of the uncertainty on the $a_{1}$ mass on the final results in this analysis several additional MC simulations were built. This included two sets of samples for each HNL mass hypothesis (and the SM scenario), where the $m_{a_{1}}$ was varied to $\pm 1 \sigma$ of the experimental average (where $\sigma = 40$\mevcc). The simulation source code was altered to use these masses instead of the average, and the $\tau$ decays were re-simulated in the same way so described in Sec.~\ref{sec:signal_samples}. For each $\pm 1\sigma$ value, sufficient statistics were generated to ensure $\sim \mathcal{O}(10^{6})$ events were reconstructed. These were then used in the final analysis to re-weight the 2D template for signal sample and entire $\tau$ MC sample. An event-by-event re-weighting was applied to the reconstructed samples and the same selection requirements were applied. The re-weighted samples were run through the same analysis code as the full unweighted data set and the limit on $|U_{\tau 4}|^{2}$ recalculated for each $\pm 1 \sigma$ mass value. The largest variation in the result will be quoted as a systematic uncertainty in Sec.~\ref{sec:results}.

\begin{figure*}[t]
\centering
\begin{tabular}{ccc}
 $m_{4}$ = 0 \mevcc & $m_{4}$ =500 \mevcc &  $m_{4}$ =1000 \mevcc \\
\includegraphics[width=2.5in,height=1.7in]{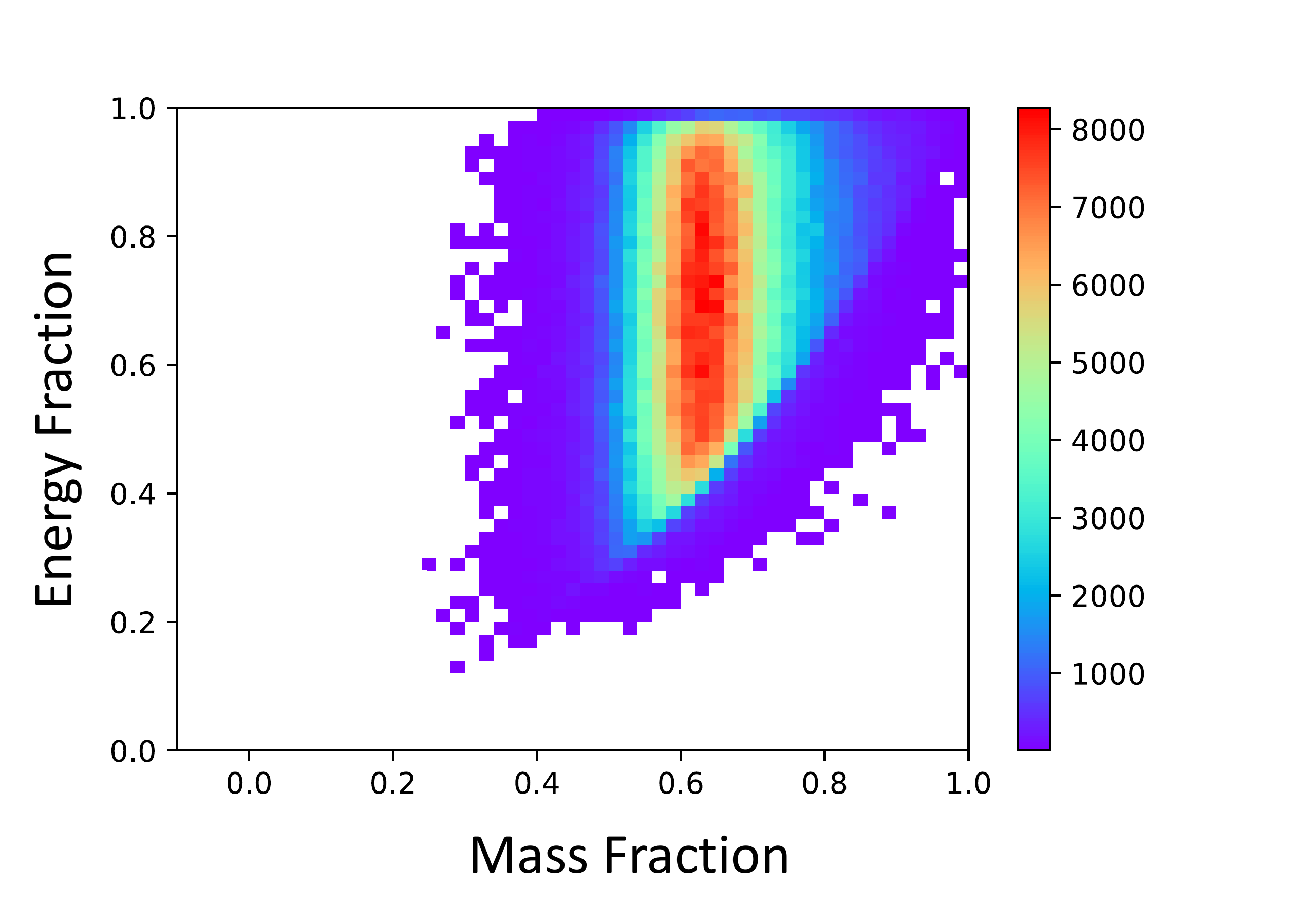} 
& \includegraphics[width=2.5in,height=1.7in]{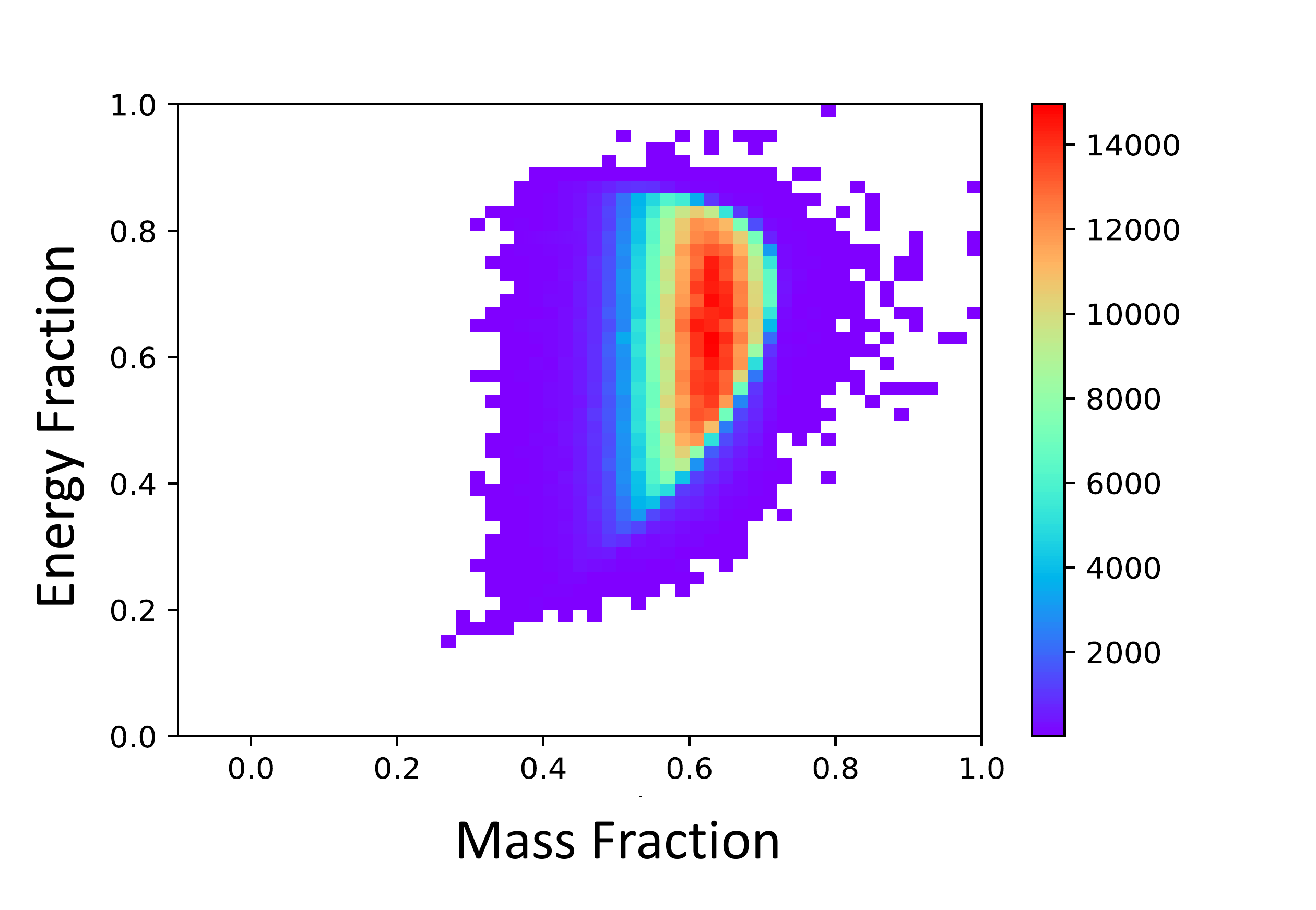} 
&\includegraphics[width=2.5in,height=1.7in]{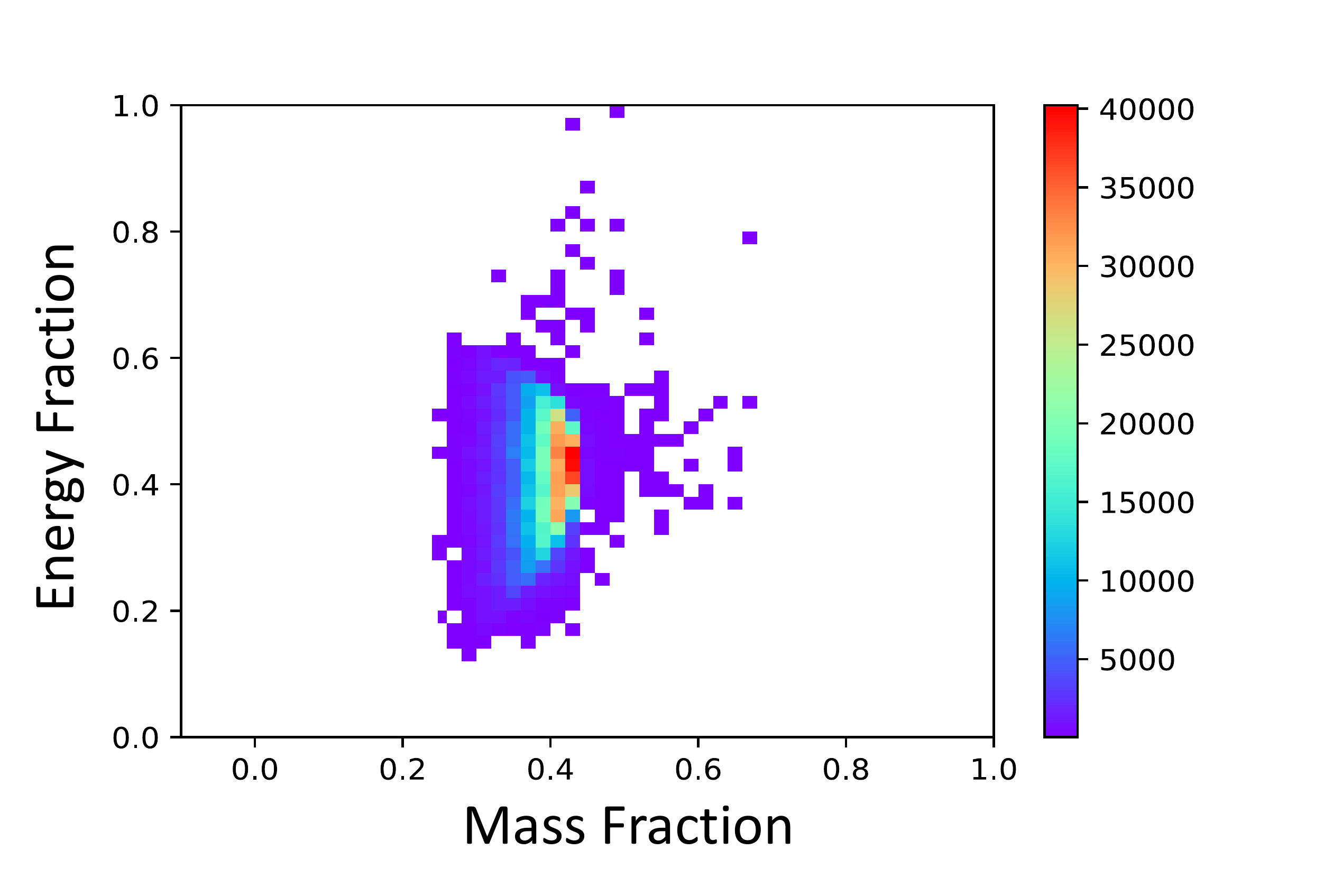}
 \\
\includegraphics[width=2.5in,height=1.7in]{figs/0MeV_1230_Templatev2.pdf} 
& \includegraphics[width=2.5in,height=1.7in]{figs/500MeV_1230_Template.pdf} 
& \includegraphics[width=2.5in,height=1.7in]{figs/1000MeV_1230_Template.pdf}
 \\
\includegraphics[width=2.5in,height=1.7in]{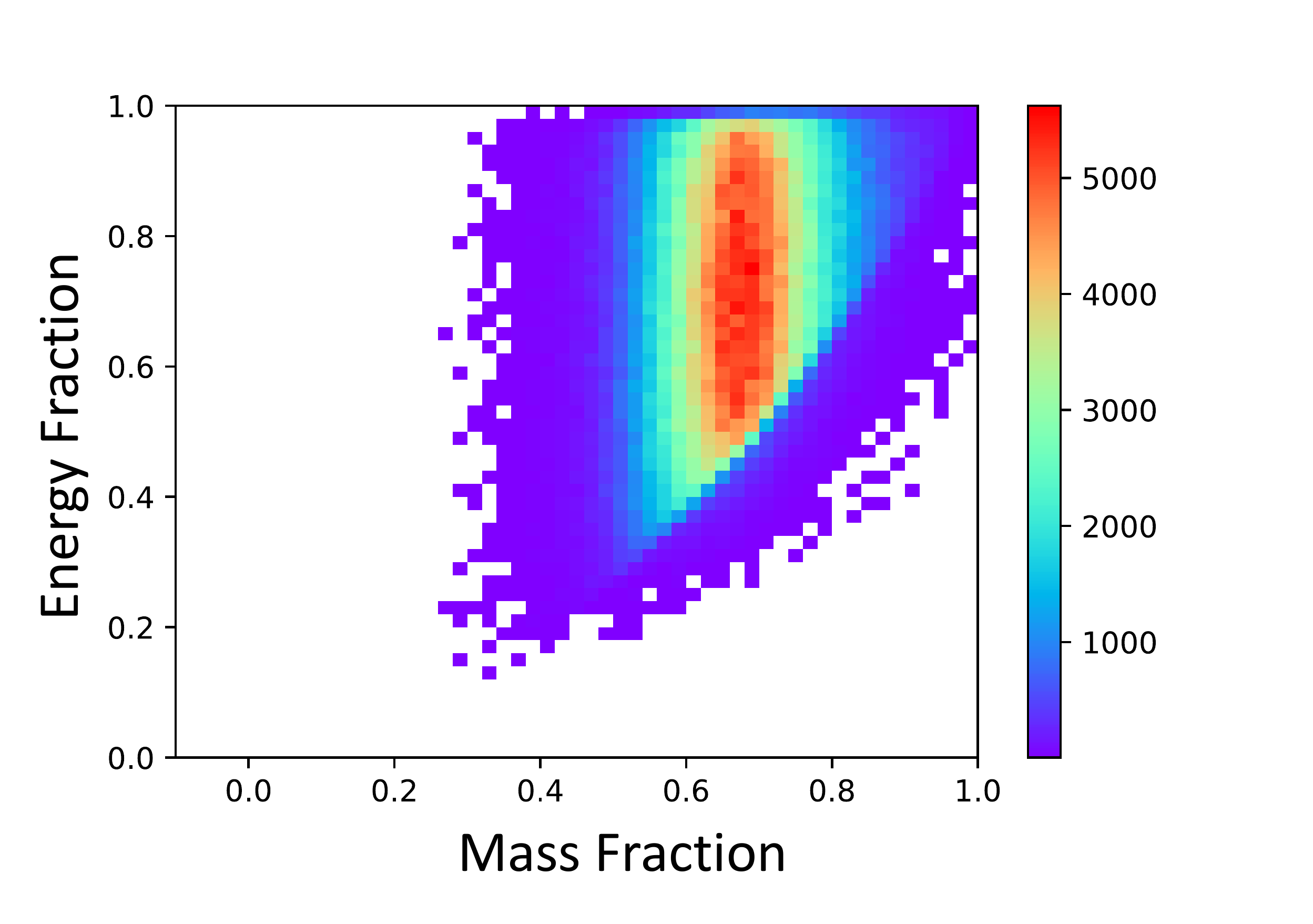} 
 & \includegraphics[width=2.5in,height=1.7in]{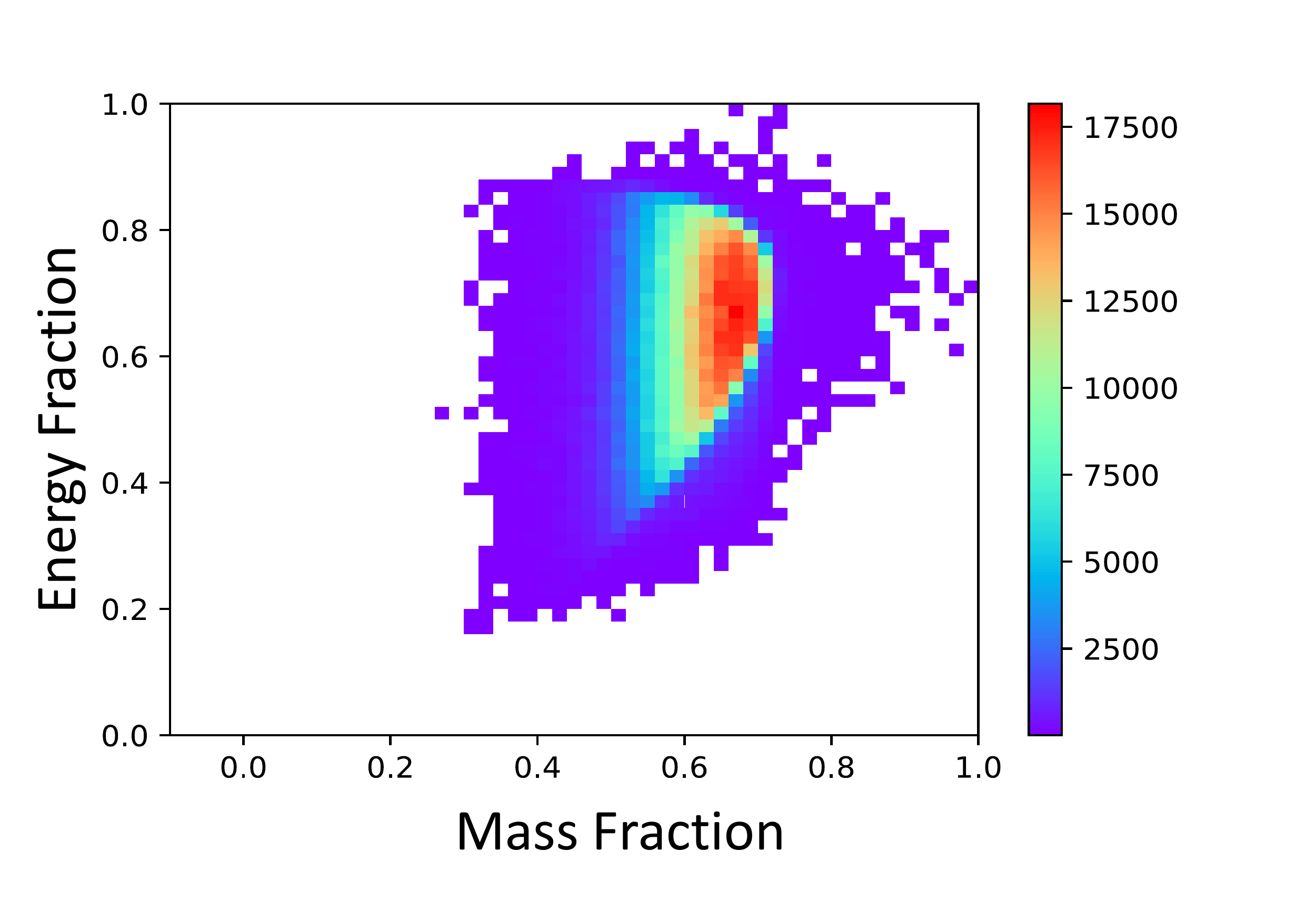} 
 &\includegraphics[width=2.5in,height=1.7in]{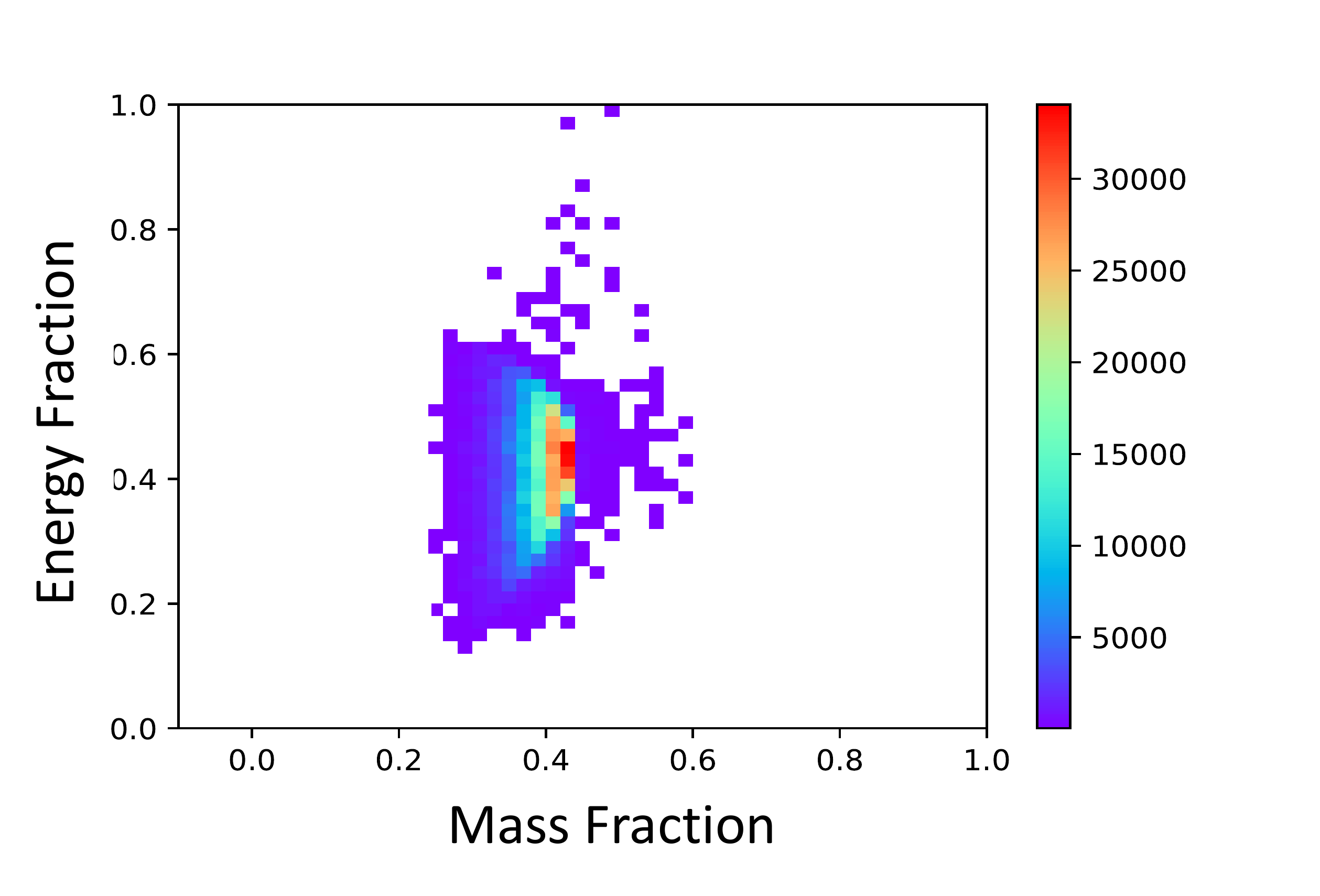}
 \\
\end{tabular}
\caption{Examples of reconstructed outgoing hadronic invariant-mass and energy ($m_{h},E_{h}$)  (as fraction of incoming $\tau$ mass and
energy) for HNL masses of $m_{4}$ = 0 \mevcc~ (left column),  500 \mevcc~ (middle column), and 1000 \mevcc~ (right column) for three $m_{a_{1}}$ possibilities: 1190 \mevcc~ (top row), 1230 \mevcc~ (middle row), and 1270 \mevcc~ (bottom row). These samples are used to re-weight the 2D template histograms and the shift in the derived value of $|U_{\tau 4}|^{2}$ is quoted as an uncertainty. }
        \label{fig:vary_A1}
\end{figure*}

Figure~\ref{fig:vary_A1} shows the example reconstructed 2D template histograms for several HNL mass hypotheses. Section~\ref{percent_a1} lists the relative shift in the means and variances of the distributions for each altered template. These relative shifts get smaller as the neutrino mass increases. For the 0 \mevcc~ scenario the mean shifts by around $\pm 2.1\%$ in $m_{h}/m_{\tau}$ and around $\pm 1.2 \%$ in $E_{h}/E_{\tau}$. For the case with $m_{4}$ = 500 \mevcc the  mean shifted by around $\pm 1.2\%$ in $m_{h}/m_{\tau}$ and around $\pm 0.5 \%$ in $E_{h}/E_{\tau}$. At $m_{4}$ = 1000 \mevcc~ the shift in $E_{h}/E_{\tau}$ disappears and the shift in $m_{h}/m_{\tau}$ is just $\pm 0.8 \%$. This is due to the allowed phase space of the hadronic system being substantially diminished as the mass of the invisible component increases. The heavier HNL takes a much larger fraction of the available energy and, therefore, the shift in the visible distributions becomes less apparent. 

A shift in reconstructed mass and energy within a bin has no effect on the end result. However, if the difference in the template is large enough it can result in events being shifted by a bin (or possibly more) relative to the distribution when the average quantities are used. Each axis has 50 bins. For the mass axis each bin represents $m_{\tau}/50 \simeq 35$ \mevcc, for the energy axis, the fraction depends on the incoming $\tau$ CM energy which has a maximum around $\sqrt{s}/2 $ GeV, meaning a bin width of $\sim$ 100 MeV. Of course, depending where the event lies within the bin, shifts in less than these values can still result in it moving bins. When the mass is shifted to the lower value the mean of the distribution in both $m_{h}$ and $E_{h}$ becomes lower, since the incoming mass-energy is lower, when the higher $m_{a_{1}}$ is used the opposite is true. The effects are symmetric about the average, $m_{a_{1}}$ = 1230 \mevcc, case.

To understand the effect of the experimental uncertainty on the value of $\Gamma_{a_{1}}$ on the end result, again additional samples were made using widths at either experimental average $\pm 1 \sigma$ value, where $\sigma$ = 35 \mevcc. The resulting shifts in the mean and RMS values of the 2D templates are relatively small ($\ll$ 1$\%$ for the both SM and signal cases). In the case of the signal samples, the relative effect gets smaller with increased outgoing HNL mass and becomes negligible for HNL masses $>$ 1000 \mevcc, in all instances the effect is small. 

It must be recognized, however, that the experimental values presented in Ref.~\cite{newpdg} span a wide range. The PDG estimates that the width is in fact somewhere between 250 - 600 \mevcc. This is a conservative band and the averaged value is contained within this estimated band. To assess the effect of this conservative uncertainty, the width was again shifted to these two values. This results in a relatively large shift of up to $\sim 6 - 7 \%$ in the $m_{h}/m_{\tau}$ RMS values, with the $E_{h}/E_{\tau}$ shifting by $\sim 1 - 3 \%$. For both axes the mean values shift by only $\sim 1 - 2 \%$  for the SM scenario. The heavier HNL signal samples are again less effected by the change.

In the final result the uncertainty on the width values will be taken from the largest change in the final result when the more conservative, estimated, widths are considered. 

\subsubsection{Uncertainties in modeling of other \texorpdfstring{$\tau$}~ channels}
\label{sec:tau_shape}

 There are two other dominant $\tau$ channels in this analysis $\tau \rightarrow 3\pi^{\pm} + \pi^{0} $ and  $\tau \rightarrow 3\pi^{\pm} + 2\pi^{0} $, with the former being, by far, the largest. These other hadronic $\tau$ decay channels involve several other intermediate states; TAUOLA models these multi-pion channels in an oversimplified way. The spectral functions involved are calculated using data. To understand the impact of uncertainty from the modeling of these modes the reconstructed samples can be re-weighted to reflect the reconstructed mass/energy obtained when the masses used for the intermediate mesons in these decay models are varied to the PDG $\pm 1 \sigma$ values. This follows the same technique as described in the previous section. 

Uncertainties in all $\tau$ hadronic models are included in the background uncertainty displayed in Fig.~\ref{fig:e_both} and Fig.~\ref{fig:mu_both}. The impact of these uncertainties in the 3-prong and other $\tau$ background shapes will be quoted in the final result as the ``modeling uncertainty."

\subsection{Systematic Uncertainty Summary Table}

Table~\ref{table:syslist} lists the relative contribution for each of the systematic uncertainties discussed in this section. 

\begin{table*}[t!]
\centering
\caption{Systematic uncertainty contribution to the event yield (in$\%$)  from each source, based on comparisons between MC simulations and data. }
\begin{tabular}{c |  c } 
 \hline 
Uncertainty &  Yield Change ($\pm$)\\ [0.5ex] 
 \hline
 \hline
Luminosity &  $0.44 \%$\\
$\sigma(ee \rightarrow \tau \tau)$ &$0.31 \%$\\
\hline
Branching Fractions (1 prong)  & e: 0.22$\%$ \\
&$\mu$: 0.22$\%$\\
\hline
Branching Fractions (3 prong)  & 3$\pi$: 0.57$\%$ \\
\hline
PID Efficiency & e: 2$\%$ \\
& $\mu $: 1$\%$ \\
&  $\pi $:  3$\%$ \\
   \hline
Bhabha Contamination &  0.2$\%$\\\
$q\bar{q}$ Contamination (data) &0.1$\%$\\
\hline
   Tracking Efficiency & negligible\\
   Detector Modeling & negligible \\
\hline
Beam Energy & negligible\\
Tau Mass & negligible\\
\hline
\end{tabular}
\label{table:syslist}
\end{table*}

\section{\label{sec:results}Results}

\begin{figure*}[t]
         \centering
         \includegraphics[width=5in]{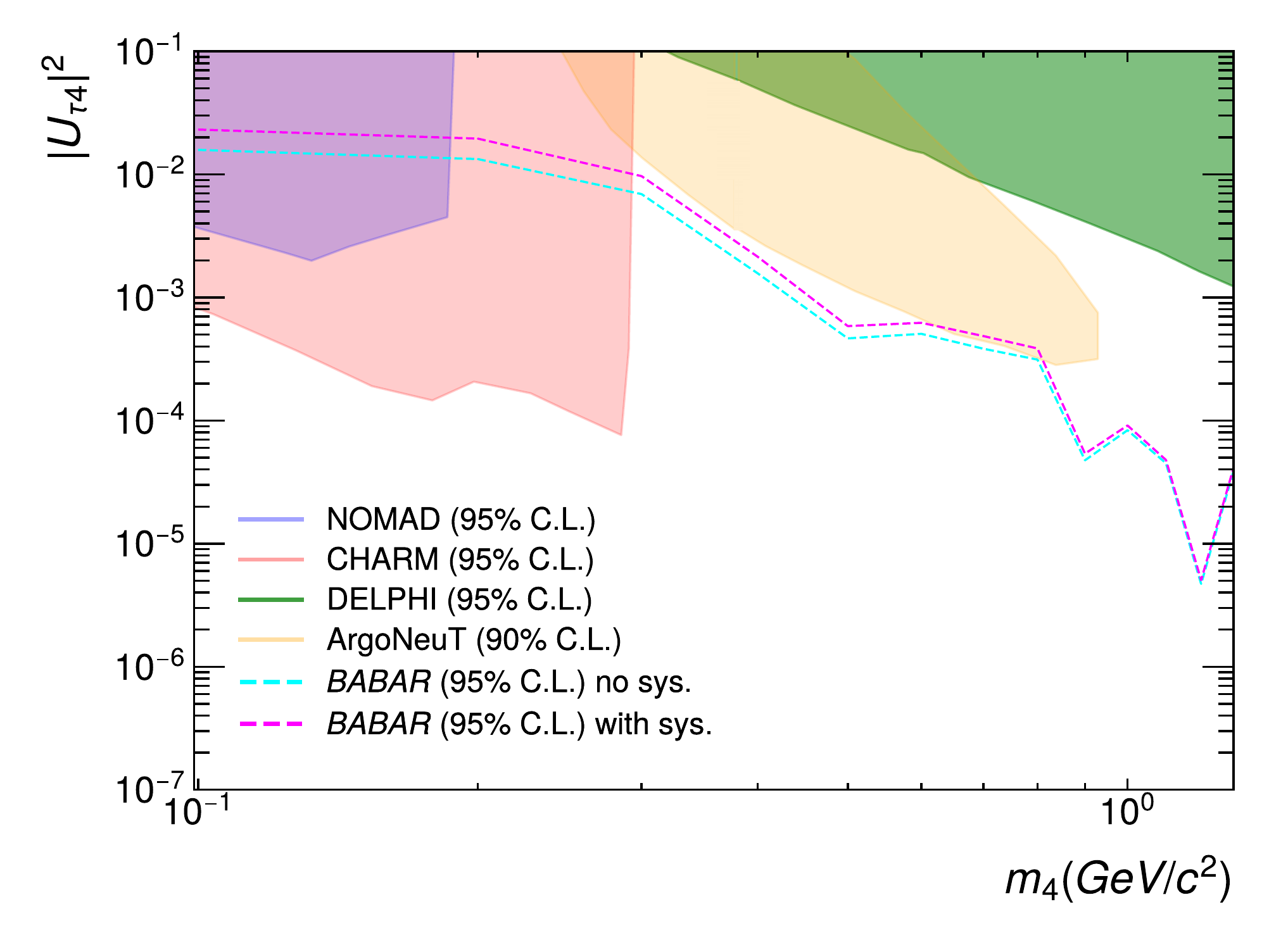}
        \caption{Upper limits at 95$\%$ C.L. on $|U_{\tau 4}|^{2}$. The magenta line represents the result when uncertainties are included. The magenta line is expected to be a very conservative upper limit. Limits from NOMAD \cite{Nomad}, CHARM \cite{CHARM} and DELPHI \cite{DELPHI} are also shown for reference. The recent ArgoNeuT result is also shown \cite{PhysRevLett.127.121801}.}
        \label{fig:limits}
\end{figure*}

For the final results, and following Eq.~\ref{eq:lik}, the data from the two channels were combined, assuming CPT symmetry holds, and the coupling to neutrinos and anti-neutrinos is identical. Figure~\ref{fig:limits} shows the upper limit at the 95$\%$ confidence level provided by this analysis using the binned likelihood technique described in Sec.~\ref{sec:likelihood}. The magenta line represents the upper limit when all systematic uncertainties are considered. To characterize deviations due to the uncertainty on $\Gamma_{a_{1}}$ the more conservative PDG estimates are used. The dominant systematic uncertainty is, by far, that due to the assumptions made within our hadronic modeling, the main contribution being uncertainty in the intermediate resonances for the $\tau$ 3-prong channel, and the dominant $\tau$ backgrounds.  Table~\ref{table:end_result} lists these 95$\%$ C.L. upper limits derived for each HNL mass hypothesis, with and without the systematic uncertainties considered. The ``with systematic uncertainty" is a conservative calculation, which includes the largest discrepancies possible in the 2D fits with consideration of experimental limits on the resonances.  The relative systematic uncertainty decreases as the mass of the hypothetical HNL increases, which is expected since it was shown in Sec.~\ref{a1} that the effects of the modeling uncertainty were less apparent at higher HNL masses.

  \begin{table*}[t]
\centering
\caption{Break down of upper limit on $|U_{\tau 4}|^{2}$ (95$\%$ C.L.) for each HNL mass hypothesis, with and without consideration of systematic uncertainty.}
\begin{tabular}{c| c | c } 
 \hline
Mass [\mevcc]  &  No Sys. &  With Sys. \\ [0.5ex] 
 \hline \hline
100 &$1.58\times 10^{-2}$&  $2.31\times10^{-2}$\\
200 &$1.33\times 10^{-2}$&  $1.95\times10^{-2}$\\
300 &$6.91\times 10^{-3}$& $9.67\times10^{-3}$\\
400 &$1.57\times 10^{-3}$& $2.14\times10^{-3}$\\
500 &$4.65\times 10^{-4}$& $5.85\times10^{-4}$\\
600 &$5.06\times 10^{-4}$& $6.22\times10^{-4}$\\
700 &$3.82\times 10^{-4}$& $4.85\times10^{-4}$\\
800 &$3.12\times 10^{-4}$& $3.85\times10^{-4}$\\
900 &$4.70\times 10^{-5}$& $5.38\times10^{-5}$\\
1000 &$8.34\times 10^{-5}$& $9.11\times10^{-5}$\\
1100 &$4.49\times 10^{-5}$& $4.78\times10^{-5}$\\
1200 &$4.70\times 10^{-6}$& $5.04\times10^{-6}$\\
1300 &$3.85\times 10^{-5}$& $4.09\times10^{-5}$\\
 \hline \hline
\end{tabular}
\label{table:end_result}
\end{table*}

\section{\label{sec:conclusion}Conclusions}

The 95 $\%$ C.L upper limits on  $|U_{\tau 4}|^{2}$ obtained in this work for  masses $100 < m_{4}<1300$ \mevcc are shown in Fig.~\ref{fig:limits}. Limits derived for the lower mass hypotheses are within the already excluded region, as expected since with this kinematic method the higher mass signals would produce the most signal/background discrimination, and, therefore, better limits. It should also be noted that the limits provided here are competitive with the current projections for experiments in the 5 $-$ 10 year time frame in this mass range including those from Belle-II, FASER and NA62 \cite{Abdullahi:2022jlv}. Looking further ahead, significant improvements are expected from the proposed facilities: the FCC-ee~\cite{FCCee} and the ILC.

\section{\label{sec:ack}Acknowledgements}

We are grateful for the extraordinary contributions of our PEP-II colleagues in achieving the excellent luminosity and machine conditions that have made this work possible. The success of this project also relies critically on the expertise and dedication of the computing organizations that support ~\babar. The collaborating institutions wish to thank SLAC for its support and the kind hospitality extended to them.

\clearpage
\bibliography{paper}

\appendix
\section{Changes in Templates with altered \texorpdfstring{$a_{1}$}~ resonance}
\label{percent_a1}

\begin{table*}[t]
\centering
\caption{Relative changes in the mean and RMS of the 2D template histograms when the $a_{1}$ mass is altered to each of its PDG $\pm 1 \sigma$ and the average value.}
\begin{tabular}{ c c  | c| c | c | c    } 
 \hline
 & &$m_{4} = 0$ \mevcc& $m_{4} = 500$ \mevcc&$m_{4} = 700$ \mevcc&$m_{4} = 1000$ \mevcc\\
$m_{a_{1}}$ [\mevcc] &&  & & \\ [0.5ex] 
\hline \hline
1190 &Mean $m_{h}/m_{\tau}$ & $-$2.09$\%$& $-$0.87$\%$ & $-$0.15$\%$&$-$0.08$\%$\\
&Mean $E_{h}/E_{\tau}$ & $-$1.14$\%$& $-$0.56$\%$ & $-$0.02$\%$& 0$\%$\\
&RMS $m_{h}/m_{\tau}$& $-$1.85$\%$ &  $-$1.51$\%$&$-$1.04$\%$&  $-$0.57$\%$\\
&RMS $E_{h}/E_{\tau}$ &$+$1.96$\%$&  $+$2.42$\%$ & $-$0.21$\%$&  $-$0.17$\%$\\

\hline

1270 &Mean $m_{h}/m_{\tau}$ &$+$2.10$\%$ & $+$0.87$\%$& $+$0.14$\%$& $+$0.07$\%$\\
&Mean $E_{h}/E_{\tau}$ &$+$1.15$\%$&  $+$0.45$\%$&$+$0.04$\%$& 0$\%$\\
&RMS $m_{h}/m_{\tau}$& $+$1.87$\%$& $+$1.57$\%$ & $+$1.23$\%$& $+$0.61$\%$\\
&RMS $E_{h}/E_{\tau}$& $-$2.06$\%$&  $-$1.72$\%$& $+$0.81$\%$& $+$0.14$\%$\\

 \hline 

\end{tabular}
\label{table:resonance}
\end{table*}

\begin{table*}[t]
\centering
\caption{Relative changes in the mean and RMS of the SM 2D template histograms (before selection requirements) when the $a_{1}$ width is altered to each of its averaged PDG $\pm 1 \sigma$, and the extremes of the PDG ``estimate".}
\begin{tabular}{ c c| c | c | c | c } 
 \hline
 & &0 \mevcc & 500 \mevcc & 700 \mevcc & 1000 \mevcc\\ [0.5ex] 
$\Gamma_{a_{1}}$ [\mevcc] & & & &\\
\hline \hline
$+$ 1$\sigma$ &Mean $m_{h}/m_{\tau}$ & $-$0.22$\%$ & $-$0.32$\%$ & $-$0.11$\%$ &$-$0.07$\%$\\
&Mean $E_{h}/E_{\tau}$ & $-$0.07$\%$ & $-$0.17$\%$ & 0.0$\%$ &0.0$\%$\\
&RMS $m_{h}/m_{\tau}$ &$+$0.40$\%$ &  $+$0.47$\%$ & $+$0.45$\%$ &$+$0.43$\%$\\
&RMS $E_{h}/E_{\tau}$ & $+$0.96$\%$ & $+$0.89$\%$ & $+$0.46$\%$ &$+$0.27$\%$\\
\hline

$-$ 1$\sigma$ &Mean $m_{h}/m_{\tau}$:& $+$0.55$\%$& $+$0.43$\%$ & $+$0.14$\%$ &$+$0.08$\%$ \\
&Mean $E_{h}/E_{\tau}$ & $+$0.28$\%$ & $+$0.17$\%$ & 0.0$\%$ &0.0$\%$\\
&RMS $m_{h}/m_{\tau}$ & $-$0.83$\%$ & $-$0.42$\%$ & $-$0.21$\%$ &$-$0.09$\%$\\
&RMS $E_{h}/E_{\tau}$ & $-$0.82$\%$ & $-$0.63$\%$ & $-$0.44$\%$ &$-$0.34$\%$\\
 \hline \hline
250 &Mean $m_{h}/m_{\tau}$:& $+$2.1$\%$& $+$1.89$\%$&$+$0.61$\%$&$+$0.55$\%$\\
&Mean $E_{h}/E_{\tau}$ &$+$1.0$\%$&$+$0.99$\%$&$+$0.27$\%$&$+$0.14$\%$\\
&RMS $m_{h}/m_{\tau}$ &$-$7.3$\%$ &$-$3.16$\%$&$-$0.89$\%$&$-$0.68$\%$\\
&RMS $E_{h}/E_{\tau}$ &$-$3.4$\%$&$-$2.98&$-$1.21$\%$&$-$0.87$\%$\\
 \hline 
600 &Mean $m_{h}/m_{\tau}$& $-$2.2$\%$& $-$1.73$\%$&$-$0.82$\%$ & $-$0.64$\%$\\
&Mean $E_{h}/E_{\tau}$ &$-$1.09$\%$&$-$1.24$\%$&$-$0.32$\%$&$-$0.06$\%$\\
&RMS $m_{h}/m_{\tau}$ &$+$6.4$\%$ &$+$3.22$\%$&$+$4.21$\%$&$+$1.72$\%$\\
&RMS $E_{h}/E_{\tau}$ &$+$2.02$\%$&$+$2.34$\%$&$+$1.26$\%$&$+$0.94$\%$\\
 \hline 
\end{tabular}
\label{table:width}
\end{table*}

\end{document}